\documentclass[11pt,longbibliography,aps,notitlepage,nofootinbib]{revtex4-1}
\usepackage{amssymb,amsmath}
\usepackage{bm}
\usepackage[pdftex]{graphicx}
\usepackage{subfig}
\usepackage{color}

\usepackage{blindtext}
\usepackage{microtype}
\usepackage{graphicx}
\usepackage{natbib}

\usepackage{siunitx,physics}
\usepackage{textcomp}
\usepackage{gensymb}
\usepackage{multirow}
\usepackage[dvipsnames]{xcolor}
\usepackage{hyperref}
\hypersetup{colorlinks=true,linkcolor=blue,filecolor=magenta,urlcolor=cyan}
\usepackage[nameinlink, noabbrev]{cleveref}
\usepackage{soul}

\usepackage[bottom]{footmisc}

\usepackage[nameinlink, noabbrev]{cleveref}

\newcommand{\beq}{\begin{equation}}
\newcommand{\eeq}{\end{equation}}

\newcommand{\MIT}{Program in Science, Technology, and Society and Department of Physics, Massachusetts Institute of Technology, Cambridge, Massachusetts 02139 USA}

\begin{document}

\title{
Tackling Loopholes in Experimental Tests of Bell's Inequality
}

\author{David I.~Kaiser}
\affiliation{\MIT}
\email{dikaiser@mit.edu }

\begin{abstract}
Bell's inequality sets a strict threshold for how strongly correlated the outcomes of measurements on two or more particles can be, if the outcomes of each measurement are independent of actions undertaken at arbitrarily distant locations. Quantum mechanics, on the other hand, predicts that measurements on particles in entangled states can be more strongly correlated than Bell's inequality would allow. Whereas experimental tests conducted over the past half-century have consistently measured violations of Bell's inequality---consistent with the predictions of quantum mechanics---the experiments have been subject to one or more ``loopholes,'' by means of which certain alternatives to quantum theory could remain consistent with the experimental results. This chapter reviews three of the most significant loopholes, often dubbed the ``locality,'' ``fair-sampling,'' and ``freedom-of-choice'' loopholes, and describes how recent experiments have addressed them. \\

Forthcoming in the {\it Oxford Handbook of the History of Interpretations of Quantum Physics}, ed. Olival Freire, Jr.~(Oxford University Press, 2021).

\end{abstract}

\date{November 18, 2020}

\maketitle

\section{Introduction}
\label{sec:Intro}

Bell's inequality \cite{Bell:1964kc,BellSpeakable} remains a hallmark achievement of modern physics, and a touchstone for efforts to distinguish between quantum mechanics and various alternatives. In particular, Bell's inequality sets a strict threshold for how strongly correlated the outcomes of measurements on two or more particles can be, if the underlying theory of nature that describes those particles' behavior satisfies certain criteria, often labeled ``local realism'' and associated with the famous paper by Albert Einstein, Boris Podolsky, and Nathan Rosen \cite{Einstein:1935rr}. In local-realist theories, the outcome of a measurement performed on a particle at one location cannot depend on actions undertaken at an arbitrarily distant location.\footnote{\small \baselineskip 10pt
Early work on Bell's inequality, including Bell's first derivation \cite{Bell:1964kc}, was deeply influenced by the EPR paper \cite{Einstein:1935rr}, in which the authors argued that particles should be considered to have definite properties on their own, prior to and independent of physicists' efforts to measure them (``realism''), and that distant events should not influence local ones arbitrarily quickly (``locality''). More recent work has clarified the minimal requirements for Bell's inequality to hold: the measurement outcome at one detector should not depend on either the detector setting or the measurement outcome at a distant detector, and the selection of detector settings on each experimental run should be independent of the properties of the particles to be measured. For recent, succinct discussions of ``local realism'' in the context of Bell's inequality, see the Appendix of Ref.~ \cite{ClauserUnspeakable2} and Section 3.1 of Ref.~\cite{MyrvoldSEP}. Note that Bell's inequality does not apply to formulations such as Bohmian mechanics \cite{Bohm:1951xw,Bohm:1951xx}, which, as Bell \cite{Bell:1964kc} noted, has a ``grossly non-local structure.''}  
Quantum mechanics is not compatible with local realism and, as Bell demonstrated, quantum mechanics predicts that measurements on pairs of particles in so-called ``entangled'' states can be {\it more strongly correlated} than the local-realist bound would allow.\footnote{\small \baselineskip 10pt For historical treatments, see Refs.~\cite{FreireOpticsLab,FreireQuantumDissidents,Bromberg,Gilder,KaiserHippies,KaiserQL,Whitaker}. For a range of philosophical responses, see Refs.~\cite{dEspagnatBook,RedheadBook,CushingBook,AlbertBook,BubInterpreting,BubBanana,BokulichBook,MaudlinBook,CramerBook,Unspeakable2,MyrvoldSEP}.
Recent popular accounts include Refs.~\cite{Orzel:2009,ZeilingerDance,Musser,Becker,Ball,Bub:2018,Greenstein,Brody}. } 

Virtually every published experimental test of Bell's inequality, stretching over half a century, has found results compatible with the predictions of quantum mechanics, and (hence) in violation of Bell's inequality.\footnote{\small \baselineskip 10pt The only published experimental test of Bell's inequality that appeared to contradict the predictions from quantum theory was published in Ref.~\cite{Faraci:1974}, though that experiment was criticized in Refs.~\cite{Kasday:1975,Clauser:1978ng}.} Yet since the earliest efforts to subject Bell's inequality to experimental test, physicists have recognized that several ``loopholes'' must be addressed before one may conclude that local-realist alternatives to quantum mechanics really have been ruled out. The loopholes consist of logical possibilities---however seemingly remote or implausible---by which a local-realist theory could give rise to correlated measurements that mimic the expectations from quantum theory, exceeding Bell's bound. (For reviews, see Refs.~\cite{Clauser:1978ng,Zeilinger:1999zz,WeihsCompendium,Brunner:2014,Larsson:2014,Giustina2017}.)

In this chapter, I discuss the three major loopholes that have been identified for experimental tests of Bell's inequality. In Section \ref{sec:CHSH}, I briefly review the form of Bell's inequality on which most experimental efforts have been focused. This form, which was introduced by John Clauser, Michael Horne, Abner Shimony, and Richard Holt \cite{Clauser:1969ny} soon after Bell published his original paper on the topic, is usually referred to as the ``Bell-CHSH inequality.'' Several of those physicists, often in close dialogue with Bell himself, were also among the first to identify various loopholes. The first of these, known as the ``locality loophole,'' is the subject of Section \ref{sec:Locality}. In Section \ref{sec:FairSampling}, I discuss the ``fair-sampling loophole,'' while in Section \ref{sec:FreedomChoice} I turn to the ``freedom-of-choice loophole.'' Brief concluding remarks follow in Section \ref{sec:conclusions}. As described below, tests of Bell's inequality have been performed on many different physical systems, subjecting different types of particles to measurements with different types of detectors. In this chapter I focus primarily on conceptual analysis of the various loopholes, more than on the details of particular experimental implementations.

\section{The Bell-CHSH Inequality and the First Experimental Tests}
\label{sec:CHSH}

\begin{figure}
    \centering
    \includegraphics[width=4in]{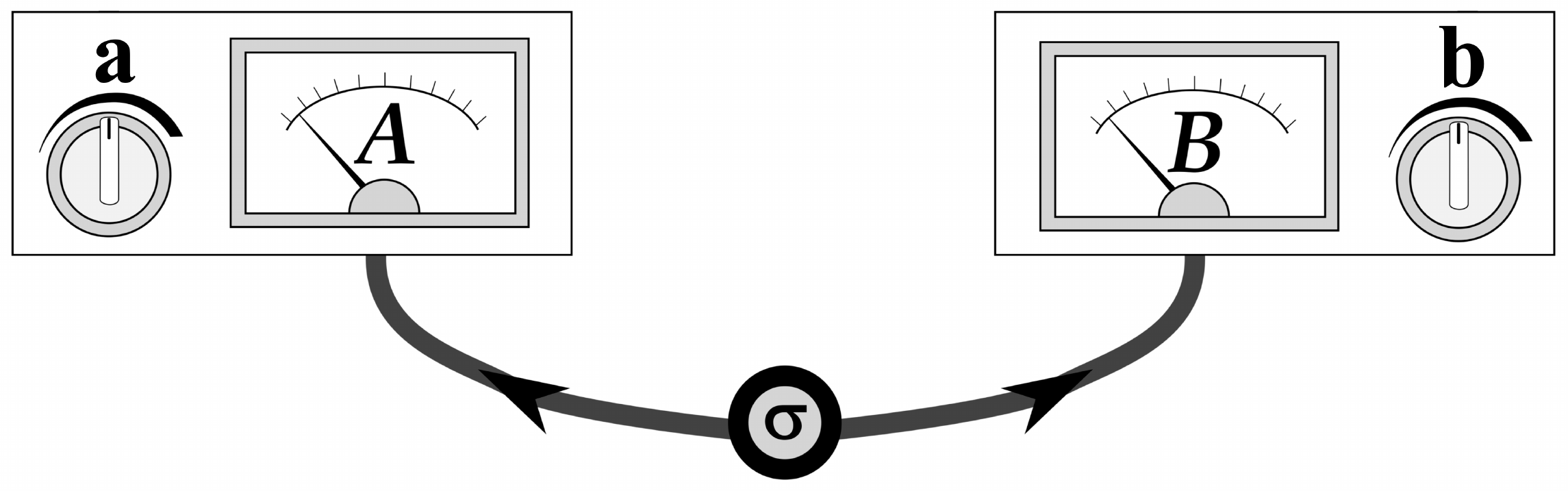}
    \caption{\small \baselineskip 10pt Schematic illustration of a typical Bell test. A source $\sigma$ emits a pair of particles, which travel in opposite directions. At each detector, a physicist selects a particular measurement to be performed by adjusting the detector settings $({\bf a}, {\bf b})$; each detector then yields a measurement outcome $(A,B)$. (Adapted from Ref.~\cite{Gallicchio:2013iva}.)}
    \label{fig:BellTest}
\end{figure}

Most experimental tests of Bell's inequality have concerned correlations among measurements on pairs of particles. Such tests can be pictured as in Fig.~\ref{fig:BellTest}: a source ($\sigma$) in the center of the experiment emits a pair of particles which travel away from the source in opposite directions. At each detector, a physicist selects a particular measurement to be performed by adjusting the detector settings (${\bf a}, {\bf b}$); each detector then yields a measurement outcome ($A, B$). For example, if the particles emitted from the source consist of pairs of electrons, a physicist at the left detector might choose to measure the spin of the left-moving electron along the ${\bf x}$-axis, or along the ${\bf y}$-axis, or along some intermediate angle; her choice of basis in which to measure the electron's spin is labeled ${\bf a}$. The physicist at the right detector chooses to measure the spin of the right-moving electron along a particular orientation in space by adjusting the detector setting ${\bf b}$. In this example, for any pair of detector settings (${\bf a}, {\bf b}$) that have been selected, the measurement outcomes $(A, B)$ at each detector can only be spin-up or spin-down. If we label the measurement outcome spin-up as $+1$ and spin-down as $-1$, then we have $A ({\bf a}), B ({\bf b}) \in \{ +1, -1\}$.\footnote{\small \baselineskip 10pt David Bohm first suggested that EPR-type experiments could be conducted using measurements of observables such as spin, which have discrete sets of possible measurement outcomes, in his influential textbook on quantum mechanics: Ref.~\cite{BohmBook}, pp.~614-622. Bell was inspired by Bohm's variation while working on Ref.~\cite{Bell:1964kc}. (See Ref.~\cite{KaiserHippies}, pp.~31-37.) On the wider impact of Bohm's textbook, see Ref.~\cite{KaiserQL}, chap.~8. A beautiful variation on Bell's original argument---which (in principle) can force an empirical contradiction between predictions from local realism and quantum mechanics with a single set of measurements rather than statistical averages over many experimental runs---concerns measurements of a discrete observable such as spin on $N$-particle entangled states, with $N \geq 3$. See Refs.~\cite{GHZ:1989,GHSZ:1990}.}  

For any pair of detector settings $({\bf a},{\bf b})$, we may construct the correlation function
\beq
E ({\bf a}, {\bf b}) \equiv \langle A ({\bf a}) \, B ({\bf b}) \rangle \, ,
\label{Edef}
\eeq
where the angular brackets indicate averages over the many experimental runs in which pairs of particles were subjected to measurements with detector settings $({\bf a},{\bf b})$. For measurements of a property such as spin, the outcomes $A ({\bf a})$ and $B ({\bf b})$ can only ever be $\pm 1$, so on any given experimental run, the product $A({\bf a}) \, B ( {\bf b})$ can only ever be $\pm 1$. Upon averaging over many runs in which the detector settings were $({\bf a}, {\bf b})$, the correlation function $E ({\bf a}, {\bf b})$ therefore satisfies $-1 \leq E ({\bf a}, {\bf b}) \leq 1$.

\begin{figure}[t]
    \centering
    \includegraphics[width=2.4in]{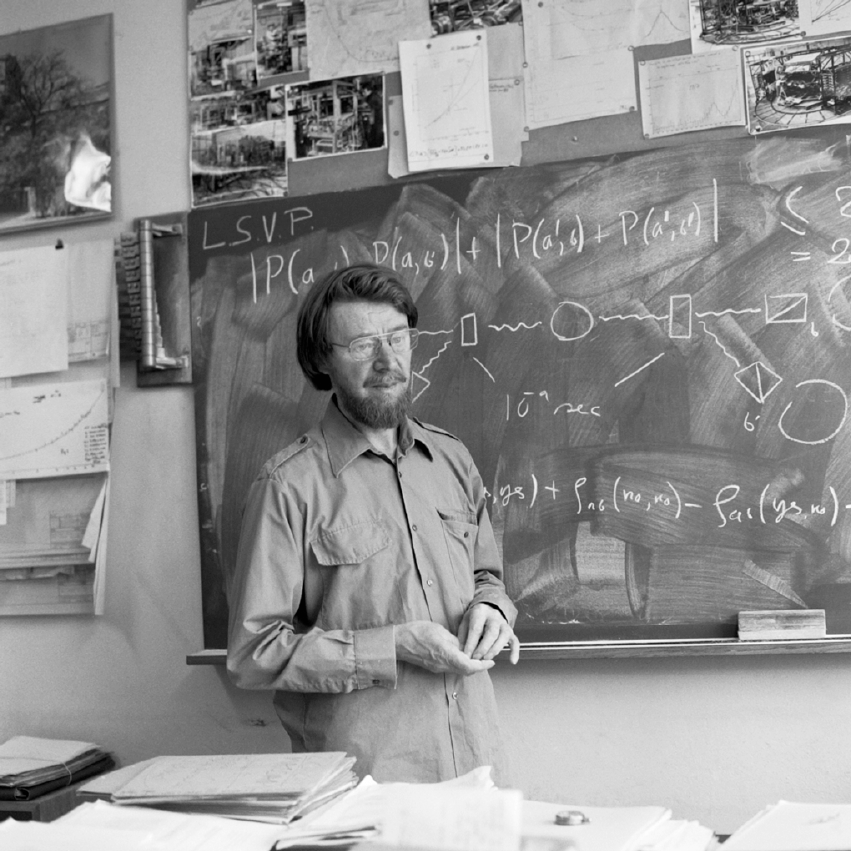}
    \caption{\small \baselineskip 10pt John S.~Bell in his office at CERN, 1982. (Courtesy CERN.)}
    \label{fig:Bell}
\end{figure}

One might try to account for the behavior of such correlation functions $E ({\bf a}, {\bf b})$ by constructing a local-realist theory and using it to calculate $p (A, B \vert {\bf a}, {\bf b})$, the conditional probability that physicists would find measurement outcomes $A$ and $B$ upon selecting detector settings ${\bf a}$ and ${\bf b}$. Bell \cite{Bell:1964kc,Bell:1971} argued that within any local-realist formulation, such conditional probabilities would take the form
\beq
\begin{split}
p (A, B \vert {\bf a}, {\bf b}) &= \int d\lambda \, p (\lambda) \, p (A, B \vert {\bf a}, {\bf b}, \lambda) \\
&= \int d \lambda \, p (\lambda) \, p (A \vert {\bf a}, \lambda) \, p (B \vert {\bf b}, \lambda) \, .
\end{split}
\label{pABab}
\eeq
Here $\lambda$ represents all the properties of the particles prepared at the source $\sigma$ that could affect the measurement outcomes $A$ and $B$. Bell imagined that whatever specific form the variables $\lambda$ took, their values on a given experimental run would be governed by some probability distribution $p (\lambda)$. Given detector setting ${\bf a}$ at the left detector, there would be some probability $p (A \vert {\bf a}, \lambda)$ to find measurement outcome $A$ at that detector, and likewise some probability $p (B \vert {\bf b}, \lambda)$ to find outcome $B$ at the right detector given detector setting ${\bf b}$.\footnote{\small \baselineskip 10pt In his original derivation, Bell \cite{Bell:1964kc} assumed that the measurement outcomes were governed by deterministic functions $A ({\bf a}, \lambda)$ and $B ({\bf b}, \lambda)$. He generalized his derivation to stochastic models, with conditional probabilities $p (A \vert {\bf a}, \lambda)$ and $p (B \vert {\bf b}, \lambda)$, in Ref.~\cite{Bell:1971}.} Note that these expressions encode ``locality'': nothing about the probability to find outcome $B$ at the right detector depends on either the setting (${\bf a}$) or the outcome ($A$) at the distant detector, and vice versa \cite{Bell:1964kc,Bell:1971,BellSpeakable}. (For a helpful and succinct discussion, see Ref.~\cite{MyrvoldSEP}.) In his original derivation, Bell \cite{Bell:1964kc} quoted from Einstein's ``Autobiographical Notes.'' As Einstein had written, ``But on one supposition we should, in my opinion, absolutely hold fast: the real factual situation of the system $S_2$ [the particle being measured at the right detector] is independent of what is done with the system $S_1$ [the particle at the left detector], which is spatially separated from the former.'' (Ref.~\cite{Einstein1949}, p.~85. See also Refs.~\cite{Howard1985,Fine1986,RennEinsteinAutoBiog}.)
See Fig.~\ref{fig:Bell}.

Bell's inequality can be cast in a particularly simple form if we consider experiments in which each particle is subjected to measurement in one of two detector settings: either ${\bf a}$ or ${\bf a}'$ at the left detector, and ${\bf b}$ or ${\bf b}'$ at the right detector. Then one may consider a particular combination of correlation functions, as one varies the pairs of detector settings:
\beq
S \equiv \big\vert E({\bf a}, {\bf b}) + E ({\bf a}', {\bf b}) - E ({\bf a}, {\bf b}') + E ({\bf a}', {\bf b}') \big\vert \, .
\label{Sdef}
\eeq
The quantity $S$, first derived by Clauser, Horne, Shimony, and Holt \cite{Clauser:1969ny}, is known as the Bell-CHSH parameter. Closely following Bell's original reasoning in Ref.~\cite{Bell:1964kc}, the CHSH authors demonstrated that for any model in which conditional probabilities $p (A, B \vert {\bf a} , {\bf b})$ took the form of Eq.~(\ref{pABab}), the parameter $S$ obeys the inequality \cite{Clauser:1969ny}
\beq
S  \leq 2 \, .
\label{CHSHineq}
\eeq
Eq.~(\ref{CHSHineq}) is known as the Bell-CHSH inequality.\footnote{\small \baselineskip 10pt As in Ref.~\cite{Bell:1964kc}, the original CHSH derivation \cite{Clauser:1969ny} applied to local-realist models in which the measurement outcomes were given by deterministic functions $A ({\bf a}, \lambda)$ and $B ({\bf b}, \lambda)$. The CHSH inequality in Eq.~(\ref{CHSHineq}) also applies to stochastic models in which $A ({\bf a}, \lambda) \rightarrow p (A \vert {\bf a}, \lambda)$ and $B ({\bf b}, \lambda) \rightarrow p (B \vert {\bf b} , \lambda)$. See, e.g., Appendix A of Ref.~\cite{Hall:2011} and Ref.~\cite{MyrvoldSEP}.} A straightforward calculation (see, e.g., Ref.~\cite{Sakurai}, pp.~223-232) suffices to show that for pairs of particles prepared in a maximally entangled state, such as 
\beq
\vert \Psi^{(\pm)} \rangle = \frac{1}{\sqrt{2}}\bigg\{ \ket{ +1 }_A \otimes \ket{ -1 }_B \pm \ket{ -1 }_A \otimes \ket{ +1 }_B \bigg\} \, ,
\label{psipm}
\eeq
quantum mechanics predicts that $S$ could {\it violate} the bound in Eq.~(\ref{CHSHineq}), achieving a maximum value
\beq
S_{QM}^{\rm max} = 2 \sqrt{2} \, .
\label{SQM}
\eeq
According to quantum mechanics, the value $S_{QM}^{\rm max}$ should arise for particular choices of settings $({\bf a}, {\bf a}')$ and $({\bf b}, {\bf b}')$. The value $S_{QM}^{\rm max} = 2 \sqrt{2}$ is known as the ``Tsirelson bound'' \cite{Tsirelson:1980}. 

Consider, for example, an experiment involving pairs of linearly polarized photons prepared in the state $\vert \Psi^{(\pm)} \rangle$ of Eq.~(\ref{psipm}), with $\ket{ +1 } \rightarrow \ket{ H }$ (horizontal polarization) and $\ket{ -1 }  \rightarrow \ket{ V }$ (vertical polarization) with respect to some orientation in space. If the photons travel along the ${\bf z}$ axis toward each detector, then the detector settings $({\bf a}, {\bf a}' , {\bf b}, {\bf b}')$ are simply unit vectors pointing at various angles within the ${\bf x}$-${\bf y}$ plane, along which polarizing filters could be oriented. A photon in state $\ket{H}_A$ with respect to orientation ${\bf a}$ would yield measurement outcome $A ({\bf a} ) = +1$, whereas a photon in state $\ket{V}_A$ along ${\bf a}$ would yield $A ({\bf a}) = -1$.\footnote{\small \baselineskip 10pt The earliest experimental tests involving polarized photons used single-channel measuring devices at each detector. Hence (most) photons whose polarization aligned with the orientation of the polarizing filter would pass through the filter and be registered by a device such as a photomultiplier tube, yielding $A ({\bf a}) = +1$, whereas photons whose polarization was perpendicular to the orientation of the polarizing filter would yield no registered detection, coded as $A ({\bf a}) = -1$. In practice, such an approach combined data on perpendicular polarization with all other reasons that the detector might have failed to register a photon on a given experimental run. Later experiments adopted two-channel measuring devices at each detector, taking advantage of the fact that a photon in state $\ket{H}_A$ with respect to orientation ${\bf a}$ will pass through the polarizing filter along a distinct trajectory than a photon in state $\ket{V}_A$. See, e.g., Refs.~\cite{AspectUnspeakable,MyrvoldSEP}.}
For this set-up, the quantum-mechanical prediction for the correlation function is simply $E ({\bf a}, {\bf b}) = - \cos (2 \theta_{\bf ab})$, where $\cos \theta_{\bf ab} \equiv \bf{a}\cdot{\bf b}$.\footnote{\small \baselineskip 10pt The form of $E ({\bf a}, {\bf b})$ in this case is easy to understand. When measured along the same orientation in space, ${\bf a} = {\bf b}$ and hence $\theta_{\bf ab} = 0^\circ$, pairs of polarized photons prepared in the state $\vert \Psi^{(+)} \rangle$ of Eq.~(\ref{psipm}) should be perfectly anti-correlated, with $E ({\bf a},{\bf b}) = -1$. If the photons in that state are measured along perpendicular orientations in space, with $\theta_{\bf ab} = 90^\circ$, then the polarization measurements should be perfectly correlated, with $E ({\bf a}, {\bf b}) = +1$. The variation of $E ({\bf a}, {\bf b})$ with $\theta_{\bf ab}$ follows from considering measurements in rotated bases. For example, if the polarizer at the left detector is rotated by angle $\varphi$, the eigenstates of the new detector setting $\tilde{\bf a}$ will be given by $\vert \tilde{H} (\varphi) \rangle_A = \cos \varphi \ket{H}_A + \sin \varphi \ket{V}_A$ and $\vert \tilde{V} (\varphi) \rangle_A = - \sin \varphi \ket{H}_A + \cos \varphi \ket{V}_A$. }
In that case, the Tsirelson bound corresponds to the choice of settings $({\bf a}, {\bf a}') = (0^\circ, 45^\circ)$ and $({\bf b}, {\bf b}') = (22.5^\circ , 67.5^\circ)$. On the other hand, both local-realist theories and quantum mechanics predict that $S \leq 2$ for choices such that ${\bf a} \cdot {\bf b} = {\bf a}' \cdot {\bf b}' = 0$ or 1, regardless of the angle between ${\bf a}$ and ${\bf a}'$.

The size of the difference between predictions from local-realist theories and from quantum mechanics---$S \leq 2$ versus $S \leq 2 \sqrt{2}$ for clever choices of detector settings---is sufficiently large that some physicists quickly began to imagine conducting experimental tests of Bell's inequality. One of the first to highlight this possibility was Henry Stapp, a research scientist at the Lawrence Berkeley Laboratory in California. Stapp had been trained in particle physics at Berkeley in the early 1950s; for his Ph.D. dissertation, he had studied spin correlations in proton-proton scattering experiments. During the summer of 1968---even before the CHSH version of Bell's inequality had appeared---Stapp wrote a preprint noting that Bell's inequality could be tested in experiments much like the proton-scattering ones on which he had previously focused. The critical update that would be required, compared to previous experiments, would be to vary the angles along which the protons' spins were measured, to avoid only measuring in bases such that ${\bf a} \cdot {\bf b} = {\bf a'} \cdot {\bf b'} = 0$ or 1. As Stapp wrote, ``The precise experiments considered here have not all actually been performed. But they are only slight variations of experiments that have been performed'' \cite{Stapp1976}.\footnote{\small \baselineskip 10pt Stapp originally composed and circulated Ref.~\cite{Stapp1976} during the summer of 1968, in advance of a conference on ``Quantum Theory and Beyond,'' which was held in Cambridge, England. The paper was not published in the conference proceedings, and Stapp later released the paper as a technical report from the Lawrence Berkeley Laboratory ``to fill continuing requests'' \cite{Stapp1976}. On Stapp's training and his early interest in Bell's inequality, see Ref.~\cite{KaiserHippies}, pp.~55-56. In 1976, around the time that Stapp circulated his technical report, M.~Lamehi-Rachti and W.~Mittig \cite{Lamehi:1976} reported results of their analysis of violations of Bell's inequality using low-energy proton-proton scattering data. They reported good agreement with the quantum-mechanical predictions, though given experimental limitations they needed to make several additional assumptions in order to put the proton scattering data into a form with which to test Bell's inequality, beyond those typically required of Bell tests \cite{Clauser:1978ng}.} 

\newpage

Around the same time, other physicists hit upon a similar idea. Abner Shimony, at the time a young professor in both the Physics and Philosophy Departments at Boston University, wondered whether data from previous correlation experiments, which had been conducted for other reasons, could be used to test Bell's inequality. Together with then-graduate student Michael Horne, he delved into the published literature, conducting what Horne playfully dubbed ``quantum archaeology.'' Much like Stapp, however, Horne and Shimony realized that previous correlation experiments had failed to consider the range of angles among the bases (${\bf a}, {\bf a}', {\bf b}, {\bf b}')$ for which quantum mechanics predicts $S > 2$.\footnote{\small \baselineskip 10pt Abner Shimony, interview with Joan Bromberg, September 9, 2002, transcript available in the Niels Bohr Library of the Center for History of Physics, American Institute of Physics, College Park, Maryland. See also Ref.~\cite{KaiserHippies}, pp.~45-46. } 

Independent of Stapp, Shimony, and Horne, John Clauser also became intrigued by the possibility of testing Bell's inequality in a laboratory experiment, while still in graduate school. He wrote directly to Bell in February 1969, asking if anyone had conducted such an experiment during the years since Bell's paper \cite{Bell:1964kc} had appeared. Upon hearing back from Bell that no one had as yet performed such an experiment---and with Bell's additional encouragement that if Clauser were to measure something different from what quantum mechanics predicts, that would ``shake the world!''---Clauser began thinking about how to perform such a test. In the midst of that work, he learned of Shimony's and Horne's interest in Bell's inequality, and soon they began to collaborate together \cite{Clauser2002}. (See also Ref.~\cite{FreireOpticsLab}, pp.~590-91, and Ref.~\cite{KaiserHippies}, pp.~43-45.)

\begin{figure}
    \centering
    \includegraphics[width=3in]{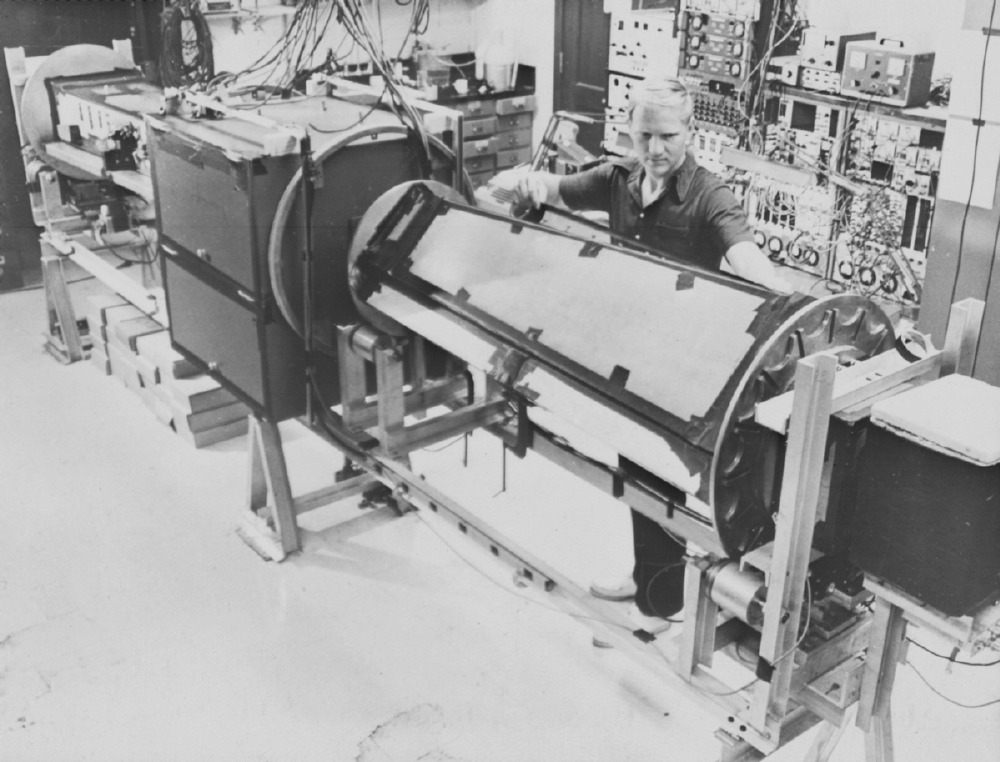}
    \caption{\small \baselineskip 10pt John Clauser working on the instrumentation with which he and Stuart Freedman conducted the first experimental test of Bell's inequality, in 1972. (Courtesy Lawrence Berkeley National Laboratory.)}
    \label{fig:Clauser}
\end{figure}

Soon after Clauser began a postdoctoral fellowship at the Lawrence Berkeley Laboratory, he asked his supervisor, quantum-electronics pioneer Charles Townes, if he could design and conduct an experimental test of Bell's inequality as a side project, separate from the main research project for which Townes had hired him. Townes agreed and arranged for Clauser to work with Stuart Freedman, at the time a graduate student at the laboratory. Freedman and Clauser used pairs of linearly polarized photons in a maximally entangled state, emitted by atomic cascades within excited calcium atoms. They mounted the polarizers in such a way that their orientations at the left and right detectors could be adjusted. The distance between the two detectors was approximately four meters. To collect their data, Freedman and Clauser first fixed the polarizer orientations at each detector, beginning with $({\bf a}, {\bf b}) = (0^\circ, 22.5^\circ)$. Then they recorded the measurement outcomes $A ({\bf a})$ and $B ({\bf b})$ within brief coincidence windows ($\Delta t = 8.1$ ns), to ensure that the pairs of measurements $(A, B)$ on a given experimental run pertained to a single pair of entangled photons that had been emitted from the source. Upon collecting many thousands of measurements on pairs of photons with the polarizers set to these orientations and averaging those results, they constructed the correlation function $E ({\bf a}, {\bf b})$. Then they paused the experiment, rotated the polarizer on the left side from orientation ${\bf a} = 0^\circ$ to ${\bf a}' = 45^\circ$ while keeping the polarizer on the right side fixed at ${\bf b} = 22.5^\circ$, and conducted new measurements with which to construct $E ({\bf a}', {\bf b})$---and so on, until they had collected sufficient data with the various joint settings $({\bf a}, {\bf b})$, $({\bf a}', {\bf b})$, $({\bf a}, {\bf b'})$, and $({\bf a'}, {\bf b}')$ to construct the combination of correlation functions needed for the Bell-CHSH parameter $S$ in Eq.~(\ref{Sdef}). Their findings were equivalent to $S = 2.388 \pm 0.072$, violating the Bell-CHSH inequality of Eq.~(\ref{CHSHineq}) by more than five standard deviations \cite{Freedman:1972zza}.\footnote{\small \baselineskip 10pt Freedman and Clauser \cite{Freedman:1972zza} reported their results in terms of a quantity $\Delta (\varphi)$, closely related to the Bell-CHSH parameter $S$ of Eq.~(\ref{Sdef}). For choices of detector settings such that $\theta_{\bf a b} = \theta_{{\bf a}' {\bf b}} = \theta_{{\bf a}'  {\bf b}'} =  \varphi$ and $\theta_{ {\bf a b}'} = 3 \varphi$, the quantities are related by $S = \vert 4 \Delta (\varphi) + 2 \vert$. They measured $\Delta (22.5^\circ) = 0.104 \pm 0.026$ and $\Delta (67.5^\circ) = -1.097 \pm 0.018$, which represent violations of $S \leq 2$ by $4.0$ and $5.4$ standard deviations, respectively. None of the early experiments found results close to saturating the (theoretical) Tsirelson bound of quantum mechanics, $S_{QM}^{\rm max} = 2 \sqrt{2} = 2.83$, largely because of limited detector efficiencies, even though they were able to measure $S > 2$ to high statistical significance. On subtleties of normalization for various Bell-like inequalities, which can complicate direct comparisons between parameters like the Bell-CHSH parameter $S$ and related quantities, see Ref.~\cite{ClauserUnspeakable2}.} (See also Refs.~\cite{Clauser:1978ng,FreireOpticsLab,KaiserHippies}.) See Fig.~\ref{fig:Clauser}.

Around the same time, Richard Holt and his graduate-school supervisor Francis Pipkin performed their own experimental test of Bell's inequality at Harvard. Like Freedman and Clauser, they conducted measurements on pairs of linearly polarized photons in a maximally entangled state, in this case using photons emitted from a particular cascade in excited mercury atoms. They used the combination of detector settings $({\bf a}, {\bf a}', {\bf b}, {\bf b}')$ predicted by quantum mechanics to yield the maximum violation of Eq.~(\ref{CHSHineq}). Yet unlike Freedman and Clauser's experiment, Holt and Pipkin measured results suggesting a strong {\it compatibility} with local-realist theories, equivalent to $S = 1.728 \pm 0.104$, easily consistent with the Bell-CHSH bound $S \leq 2$ and in strong disagreement with the prediction from quantum mechanics. They circulated a preprint of their results but never pursued formal publication, given their own lingering doubts about possible systematic errors. (See Ref.~\cite{FreireOpticsLab}, p.~595.) Three years later, Clauser repeated their experiment, using the same cascade within excited mercury atoms to produce the entangled photons, and measured the equivalent of $S = 2.308 \pm 0.0744$, a violation of the Bell-CHSH inequality by more than 4 standard deviations. (Around the same time, Edward Fry and Randall Thompson independently performed a Bell test at Texas A \& M University using entangled photons from excited mercury atoms, and, like Clauser, measured a strong violation of Bell's inequality; see Refs.~\cite{FryThompson1976,Clauser:1978ng,FryUnspeakable}.) In the course of repeating the Holt-Pipkin experiment, Clauser found that the measured correlations depended sensitively upon stresses both in the walls of the glass bulb containing the mercury vapor as well as in the lenses used to focus the emitted photons toward their respective detectors. Although Holt and Pipkin had reported observing similar stresses in the mercury bulb during their experiment, they had not attempted to repeat their measurements \cite{Clauser:1976,Clauser:1978ng,FreireOpticsLab}.\footnote{\small \baselineskip 10pt Holt and Pipkin actually measured a closely related quantity to the Bell-CHSH parameter $S$, which had been introduced in Ref.~\cite{Freedman:1972zza}: $\vert R (\varphi) - R (3 \varphi ) \vert/ R_0$, where $R (\varphi)$ is the number of coincidence count rates when the relative orientation of the polarizers at the two detectors is equal to $\theta_{\bf ab} = \varphi$, and $R_0$ is the coincidence count rate when both polarizers are removed from their respective detectors. According to quantum mechanics, for ideal measurements the parameter $R (\varphi) = {1 \over 4} [1 + \cos (2 \varphi ) ]$. For $\varphi = 22.5^\circ$, the quantum-mechanical prediction is therefore $\vert R (\varphi ) - R (3 \varphi) \vert / R_0 = \sqrt{2} / 4$, whereas local-realist models predict $\vert R (\varphi) - R (3 \varphi) \vert / R_0 \leq 1/4$. Holt and Pipkin reported $\vert R (22.5^\circ ) - R (67.5^\circ )\vert / R_0 = 0.216 \pm 0.013$, considerably below the local-realist threshold of $0.25$, much less the quantum-mechanical prediction of $0.35$ \cite{Clauser:1978ng}. When Clauser repeated their experiment \cite{Clauser:1976}, he measured $\vert R (22.5^\circ) - R (67.5^\circ) \vert / R_0 = 0.2885 \pm 0.0093$, violating the local-realist bound by $4.1$ standard deviations.}  

Since Freedman and Clauser's original experiment \cite{Freedman:1972zza}, virtually every published test has measured violations of Bell's inequality, consistent with the quantum-mechanical predictions \cite{Clauser:1978ng,Brunner:2014}. One might therefore ask: following the 1976 repetitions of the Holt-Pipkin experiment \cite{Clauser:1976,FryThompson1976}, alongside similar experiments completed during the mid-1970s \cite{Clauser:1978ng,FreireOpticsLab}, why have physicists continued to subject Bell's inequality to experimental test? The answer is that each of the early experiments was subject to multiple loopholes: explanations consistent with local realism that could (in principle) account for the experimental results.

\section{The Locality Loophole}
\label{sec:Locality}

The first loophole that physicists identified for tests of Bell's inequality is often referred to as the ``locality loophole.'' It concerns the flow of information during a given experimental run. In particular, could any communication among elements of the experiment---with information traveling at or below the speed of light---account for the strong correlation among measurement outcomes, even if the particles being subjected to test really did obey local realism?

\begin{figure}
    \centering
    \includegraphics[width=4.9in]{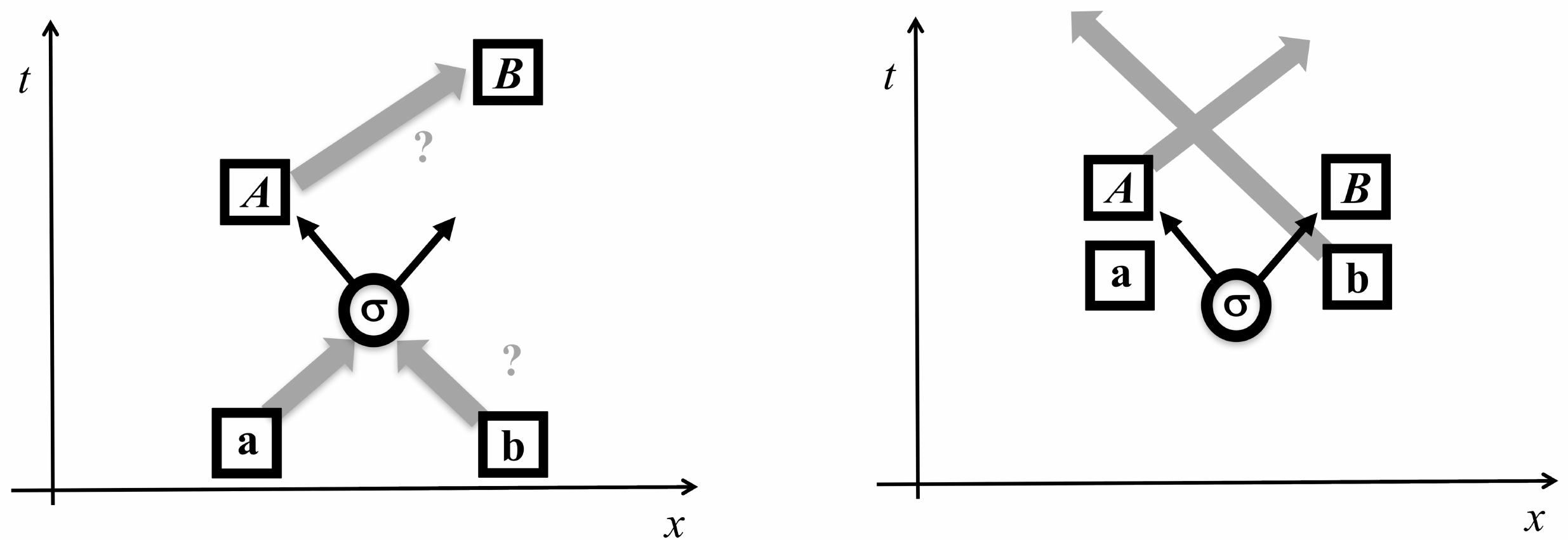}
    \caption{\small \baselineskip 10pt During a test of the Bell-CHSH inequality, experimenters must select the detector settings ${\bf a}$ and ${\bf b}$; emit the pair of entangled particles from the source $\sigma$; and perform a measurement on each particle, yielding outcomes $A$ and $B$. In the space-time arrangement shown on the left, information traveling at or below the speed of light (thick gray lines) could account for the correlations among measurement outcomes even if the particles obeyed local realism. In the space-time arrangement shown on the right, the relevant events have been spacelike separated, so that such information exchange would not suffice to account for the measured correlations. }
    \label{fig:locality}
\end{figure}

Each experimental run in a test of the Bell-CHSH inequality involves (at least) five relevant events, whose space-time arrangement we may depict as in Fig.~\ref{fig:locality}: experimenters must select the detector settings ${\bf a}$ and ${\bf b}$, emit the entangled particles from the source $\sigma$, and perform a measurement on each particle, yielding outcomes $A$ and $B$. (As usual, we adopt coordinates such that light travels one unit of space in one unit of time, so that light-like trajectories follow $45^\circ$ diagonals.) If the experimenters {\it first} select the detector settings and {\it later} emit particles from the source (as shown on the left side of Fig.~\ref{fig:locality}), then a local-realist description could readily account for the observed correlations between $A$ and $B$. In such a case, the measurement outcome $A$ could depend on information about detector setting ${\bf b}$ and/or outcome $B$ could depend on setting ${\bf a}$, such that the expression in the integrand of the top line of Eq.~(\ref{pABab}) would no longer factorize: $p (A, B \vert {\bf a}, {\bf b}, \lambda) \neq p (A \vert {\bf a}, \lambda) \, p (B \vert {\bf b}, \lambda)$.
Likewise, if the measurement on the left-moving particle were completed well before the measurement on the right-moving particle, then the detector on the right side could exploit information about the distant measurement to arrange a correlated outcome. Such explanations would be compatible with local realism.
On the other hand, if special care were taken with the space-time arrangement of these five crucial events, as shown on the right side of Fig.~\ref{fig:locality}, then the locality assumptions under which the bottom line of Eq.~(\ref{pABab}) had been derived would hold, and none of the scenarios depicted on the left side of Fig.~\ref{fig:locality} would be available for a local-realist account of the measured correlations.\footnote{\small \baselineskip 10pt Several authors have considered {\it retrocausal} models, in which $p (A, B \vert {\bf a}, {\bf b}, \lambda) \neq p (A \vert {\bf a}, \lambda) \, p (B \vert {\bf b}, \lambda)$ because of the flow of certain information backwards in time from the future light cone of a given experimental run. See Refs.~\cite{Beauregard:1978,Argaman:2010,PriceWharton:2015}. }

John Bell articulated a version of the locality loophole in his original article on Bell's inequality \cite{Bell:1964kc}. He closed his now-famous paper by writing that the quantum-mechanical predictions might apply ``only to experiments in which the settings of the instruments are made sufficiently in advance to allow them to reach some mutual rapport by exchange of signals with velocity less than or equal to that of light.'' It would therefore be ``crucial,'' he concluded, to conduct experiments ``in which the settings are changed during the flight of the particles.''\footnote{\small \baselineskip 10pt Bell \cite{Bell:1964kc} cited an article by David Bohm and Yakir Aharonov \cite{BohmAharonov}, in which they had argued that a proper test of the original Einstein-Podolsky-Rosen scenario should include the provision to change the detector settings while the particles were ``still in flight.'' Bohm had made a similar observation in his discussion of EPR in his 1951 textbook: Ref.~\cite{BohmBook}, p.~622.}
John Clauser agreed. In his first letter to Bell (written in February 1969), Clauser noted that ``it might also be possible to `rotate' the polarizers by means of magneto-optic effects while the photons are in flight'' (quoted in Ref.~\cite{FreireOpticsLab}, p.~591). Achieving such fast switching among polarizer orientations, however, proved to be quite a technical challenge. As noted in Section \ref{sec:CHSH}, when Freedman and Clauser conducted their experiment in 1972, they first manually set ${\bf a}$ and ${\bf b}$ for a given run by adjusting their polarizers to particular orientations before emitting the entangled photons. (Holt and Pipkin followed the same approach in their 1973 test, as did the other experiments conducted during the 1970s; see Refs.~{Clauser:1978ng,FreireOpticsLab}.) Bell returned to the point when summarizing discussions at a 1976 summer school devoted to the foundations of quantum mechanics, declaring that the experiments that had been conducted to date ``have nothing to do with Einstein locality,'' and that a test ``of the very highest interest'' would be one in which ``the polarization analyzers are in effect re-set while the photons are in flight'' (quoted in Ref.~\cite{FreireOpticsLab}, p.~606).

A participant in that 1976 summer school, Alain Aspect, was already hard at work on just such an experiment \cite{Aspect:1976}. Together with colleagues Jean Dalibard and G\'{e}rard Roger at the Institut d'Optique Th\'{e}orique et Appliqu\'{e}e in Orsay, near Paris, Aspect designed and built an experiment with fast-changing acoustico-optical switches inserted in the photons' paths from the source $\sigma$ to the left and right detectors. Depending on which orientation the optical switch on the left side happened to be in when a photon arrived, that photon would be directed toward one of two polarizing filters, oriented at different angles; and likewise for the optical switch on the right side. The polarizer orientations $({\bf a}, {\bf a}')$ on the left and $({\bf b}, {\bf b}')$ on the right were fixed in advance, while the optical switches on the left and right sides changed every 10 ns. The detectors in Aspect's experiment were each about 6 meters from the source of entangled particles, which meant that photons emitted from the source would require at least 20 ns to travel to their respective detectors. Hence the optical switches on each side changed one or more times while each pair of photons was in flight. The particular detector settings $({\bf a}, {\bf a}')$ and $({\bf b}, {\bf b}')$ that each photon encountered therefore had not been fixed at the time of the photons' emission, and the measurements at each detector were completed such that no signal traveling at light speed could inform the distant detector about the settings or outcome at the local detector before each side had completed its measurement. Aspect and his colleagues thus performed the first Bell test, in 1982, in which the critical events were arranged as in the right side of Fig.~\ref{fig:locality}. Even with this improved space-time arrangement, they measured a violation of the Bell-CHSH inequality by five standard deviations \cite{Aspect:1982fx}. (See also Refs.~\cite{AspectUnspeakable,FreireOpticsLab,KaiserHippies}.)

Aspect and his colleagues noted in their original article that their acoustico-optical switches did not determine the detector settings $({\bf a}, {\bf a}')$ or $({\bf b}, {\bf b}')$ in a truly random manner. Instead, the switches operated quasi-periodically, which suggested---at least in principle---that information would have been available at the photons' source $\sigma$, in advance of each emission event, that could have sufficed to predict the detector settings that each photon would ultimately encounter  \cite{Aspect:1982fx,Zeilinger:1986}. In addition, the detectors on the left and right sides in Aspect's experiment were linked in real time via electronic coincidence circuits; results from each side were not recorded independently \cite{Aspect:1982fx}. Each of these points suggested that although Aspect's 1982 experiment clearly represented an enormous milestone in tests of Bell's inequality, the ``locality'' loophole had not yet been closed conclusively.

More than fifteen years later, Anton Zeilinger and his group, at the time based at the University of Innsbruck in Austria, completed a new Bell test that addressed the locality loophole head on. Led by Gregor Weihs, the group set up two detector stations 400 meters apart, making it easier to determine the measurement outcomes $A$ and $B$ at spacelike-separated events. (For this experiment, the light travel time between detectors was $1.3 \, \mu$s rather than the 40 ns in Aspect's experiment in Orsay.) Each of the detector stations was equipped with its own quantum random number generator (QRNG), a device that could output a fresh bit (either a 0 or a 1) at a rate of 500 MHz.\footnote{\small \baselineskip 10pt The quantum random number generator (QRNG) used in Ref.~\cite{Weihs:1998} produced a rapid bitstream of 0's and 1's by shining the output from a light-emitting diode onto a beam splitter. In principle, each photon encountering the beam splitter had a 50-50 chance to be transmitted or reflected. Each path (transmission and reflection) was monitored by a photomultiplier capable of detecting single photons. Depending on which detector recorded a photon within a very brief time interval ($\Delta t \simeq 2$ ns), the device would output a 0 or a 1. } The output from each random-number generator was linked to an electro-optical modulator, a device that could quickly rotate the basis in which a photon's polarization would be measured by an angle proportional to the applied voltage, changing bases at a frequency up to 30 MHz. Each detector station also had its own atomic clock with an accuracy of $0.5$ ns, with which the time of each detection event at each detector could be recorded. Using this scheme, information about the detector settings $({\bf a}, {\bf a}')$ and $({\bf b}, {\bf b}')$ for a given run should not have been available at either the emission event $\sigma$ or at the distant measurement events that yielded $A$ or $B$, and no direct link connected the two detector stations, more conclusively achieving the space-time arrangement depicted on the right side of Fig.~\ref{fig:locality}. The group measured $S = 2.73 \pm 0.02$, a violation of the Bell-CHSH inequality by more than 35 standard deviations \cite{Weihs:1998}. (See also Refs.~\cite{WeihsUnspeakable,Gisin:1998,Aspect:1999}.)

\section{The Fair-Sampling Loophole}
\label{sec:FairSampling}

Near the conclusion of their article, Weihs, Zeilinger, and their colleagues noted that ``while our results confirm the quantum theoretical predictions, we admit that, however unlikely, local realistic or semiclassical interpretations [of their experimental results] are still possible,'' if one invoked a different loophole than locality: ``we would then have to assume that the sample of pairs registered is not a faithful representative of the whole ensemble emitted'' \cite{Weihs:1998}. After all, in their experiment, they had successfully completed measurements on only about $5\%$ of all the photon pairs that had been emitted from the source. Tests of the Bell-CHSH inequality require performing statistical averages over measurements on many pairs of particles. What if the subset of particles that was successfully detected had been drawn from some biased sample, skewing the statistical results? This second loophole has been dubbed the ``detector-efficiency loophole'' or ``fair-sampling loophole.''

Physicists typically define the ``efficiency,'' $\eta$, of a given detector as the probability that for any particle impinging upon the device, the detector will register a definite measurement outcome. In any real experiment, the efficiency will be less than $100\%$, with $0 < \eta < 1$. If one assumes that the detector efficiencies are identical for the two detector stations in a test of the Bell-CHSH inequality, and that the detectors operate independently of each other, then only in a fraction $\eta^2$ of experimental runs will each detector successfully perform a measurement on its member of an entangled pair. In a fraction $\eta (1 - \eta)$ of experimental runs, only the left detector will register a measurement of its particle while the right detector registers nothing, and in a separate fraction $\eta (1 - \eta)$ of runs, only the right detector will complete a measurement. Finally, in a fraction $(1 - \eta)^2$ of runs, neither detector will register a measurement. If one only considers experimental runs in which at least one detector completes a measurement, then the Bell-CHSH inequality of Eq.~(\ref{CHSHineq}) is modified to read  
\beq
S \leq \frac{ 4}{\eta} - 2 \, ,
\label{Seta}
\eeq
with the Bell-CHSH parameter $S$ defined in Eq.~(\ref{Sdef}).\footnote{\small \baselineskip 10pt If one assumes perfect detector efficiencies, $\eta \rightarrow 1$, then the correlation function $E ({\bf a}, {\bf b})$ in Eq.~(\ref{Edef}) may be evaluated as $E ({\bf a}, {\bf b}) = \sum_{A,B = \pm 1} ( A B \, N_{\bf ab}^{AB} ) / N_{\bf ab}^{\rm tot}$, where $N_{\bf ab}^{\rm tot}$ is the total number of entangled pairs that are emitted when the detector settings are $({\bf a}, {\bf b})$, and $N_{\bf ab}^{AB}$ is the number of double-coincidence measurements in which the left and right detectors yield $\{ A, B \} \in \{ +1, -1 \}$. However, if one takes into account imperfect detector efficiencies, one may define $E' ({\bf a}, {\bf b}) = \sum_{A, B = \pm 1, 0} (A B \, N_{\bf ab}^{AB} ) / N_{\bf ab}^{AB}$, in which $A = 0$ ($B = 0$) indicates the lack of a successful measurement at the left (right) detector. If one neglects the (unobservable) runs in which neither detector completes a measurement, then one finds $E' ( {\bf a}, {\bf b}) = [ N_{\bf ab}^{\rm double} / (N_{\bf ab}^{\rm double} + N_{\bf ab}^{\rm single} ) ] E ({\bf a}, {\bf b}) = [ \eta / (2 - \eta) ] E ({\bf a}, {\bf b} )$, where the number of double-coincidence measurements is $N_{\bf ab}^{\rm double} = \eta^2 \, N_{\bf ab}^{\rm tot}$ and the number of single-sided measurements is $N_{\bf ab}^{\rm single} = 2 \eta (1 - \eta) N_{\bf ab}^{\rm tot}$. One may then construct $S' \equiv \vert E' ({\bf a}, {\bf b}) + E' ({\bf a}', {\bf b}) - E' ({\bf a}, {\bf b}') + E' ({\bf a}', {\bf b}')\vert $, and, using the same arguments as in Ref.~\cite{Bell:1971}, derive that $S' \leq 2$ for any local-realist model in which $p (A, B \vert {\bf a}, {\bf b}$) takes the form of Eq.~(\ref{pABab}). Taking into account the different normalizations of $E' ({\bf a}, {\bf b})$ and $E ({\bf a}, {\bf b})$, one then arrives at the updated inequality for the original Bell-CHSH parameter $S$, defined in terms of $E ({\bf a}, {\bf b})$, as in Eq.~(\ref{Seta}).  } (See Refs.~\cite{Clauser:1974,GargMermin:1987,Eberhard:1993,WeihsCompendium}.) In the limiting case of detectors with perfect efficiency, $\eta \rightarrow 1$, Eq.~(\ref{Seta}) reverts to the original form of the Bell-CHSH inequality, $S \leq 2$. On the other hand, if the efficiency of each detector is below a critical threshold, $\eta \leq \eta_*$ with 
\beq
\eta_* \equiv 2 \left( \sqrt{2} - 1 \right) = 0.828 \, ,
\label{etastar}
\eeq
then the inequality in Eq.~(\ref{Seta}) becomes $S \leq 2 \sqrt{2}$, indistinguishable from the Tsirelson bound for quantum mechanics, $S_{QM}^{\rm max}$ of Eq.~(\ref{SQM}). In other words, if the detector efficiencies are less than $82.8\%$, then a local-realist explanation could account for any experimental result that found $2 \leq S \leq 2 \sqrt{2}$ simply by invoking the argument that the pairs of particles that happened to be detected during the experiment represented a biased (rather than ``fair'') sample of all the pairs that had been emitted. Several physicists developed explicit local-realist models, with conditional probabilities $p (A,B \vert {\bf a}, {\bf b})$ satisfying Bell's form in Eq.~(\ref{pABab}), that could exploit non-detection events at either detector in order to mimic the predictions from quantum mechanics. (See Refs.~\cite{Pearle:1970,Clauser:1974,Fine:1982,Marshall:1983,GargMermin:1987,Eberhard:1993}.) 

Although the fair-sampling loophole was identified as early as 1970 \cite{Pearle:1970}---and Clauser, Horne, and Shimony focused on it in various papers during the 1970s \cite{Clauser:1974,Clauser:1978ng}---addressing this loophole in a real experiment proved to be quite challenging, given technological limitations on available instrumentation. In fact, more than thirty years elapsed between the  identification of the loophole and the earliest experiments to address it. The first groups to attempt Bell tests that closed the fair-sampling loophole used pairs of slow-moving, entangled ions in high-fidelity magnetic traps, rather than entangled photons. Since the traps kept the ions accessible for long periods of time, the teams achieved very high efficiencies, $\eta \simeq 0.98$, easily above the critical threshold $\eta_*$, and managed to measure violations of the Bell-CHSH inequality. A group at the U.S. National Institute of Standards and Technology (NIST) in Boulder, Colorado performed such a test on pairs of beryllium ions in 2001, finding $S = 2.25 \pm 0.03$ \cite{Rowe:2001}, and a separate group, based at the University of Maryland, measured $S = 2.22 \pm 0.07$ with pairs of ytterbium ions in 2008 \cite{Matsukevich:2008}. (See also Ref.~\cite{Ansmann:2009}.)

\newpage

A few years later, in 2013, two groups exploited advances in highly efficient single-photon detectors to conduct Bell tests with polarization-entangled photons that closed the fair-sampling loophole. By using cryogenically cooled transition-edge sensors (TES) operating at the superconducting transition, the teams achieved detector efficiencies $\eta \geq 0.90$. By conducting short-distance tests in which the entangled photons traveled to their respective detectors via carefully shielded optical fibers, the total losses between emission at the source and measurements at the detectors remained sufficiently low to enable tests of a Bell-type inequality while closing the fair-sampling loophole.\footnote{\small \baselineskip 10pt The photon experiments in Refs.~\cite{Christensen2013,Giustina2013} each measured violations of a Bell-type inequality first derived by Philippe Eberhard \cite{Eberhard:1993}, who demonstrated that the fair-sampling loophole could be closed with overall efficiencies {\it below} $\eta_*$ in Eq.~(\ref{etastar}) if one performed measurements on pairs of particles in the {\it non}-maximally entangled state $\vert \Psi \rangle = [1 + r^2 ]^{-1/2} \left( \ket{H}_A \otimes \ket{V}_B + r \ket{V}_A \otimes \ket{H}_B \right)$, with $r$ a real parameter within the range $0 < r < 1$. In addition, these experiments exploited recent advances to efficiently produce polarization-entangled photons via spontaneous parametric down conversion \cite{Kwiat:1995}: photons of a particular frequency are directed from a pump laser onto a special nonlinear crystal, which absorbs the incoming photons and emits pairs of photons that conserve overall energy, linear momentum, and angular momentum.} 
One group measured violations of the inequality by nearly 8 standard deviations \cite{Christensen2013}, and the other group by more than 65 standard deviations \cite{Giustina2013}. As an indication of the technical challenges posed by these experiments, note that each of them---those using trapped ions in the early and mid-2000s \cite{Rowe:2001,Matsukevich:2008} and those using entangled photons in 2013 \cite{Christensen2013,Giustina2013}---focused {\it only} on closing the fair-sampling loophole, and did not even attempt to address the locality loophole. 

An enormous milestone was achieved late in 2015, when three groups performed Bell tests that closed both the locality and fair-sampling loopholes in the same experiments. The first group to accomplish this feat was directed by Ronald Hanson at the Delft University of Technology in the Netherlands. Hanson's group set up three stations across the university campus. Stations {\it A} and {\it B}, which were separated by $1.3$ km, each included a single electron spin degree of freedom, associated with a single nitrogen-vacancy (NV) defect center in a diamond chip. Each spin was entangled with a single photon; the photons from stations {\it A} and {\it B} were then transmitted to a central station {\it C} via optical fibers. Upon arrival at station {\it C}, the photons from stations {\it A} and {\it B} were subjected to a Bell-state measurement, which (as in entanglement-swapping protocols \cite{Zukowski:1993}) projected the associated electron spins at stations {\it A} and {\it B} into a maximally entangled state. After the heralding photons left stations {\it A} and {\it B} but before they arrived at station {\it C}, quantum random number generators (QRNGs) at stations {\it A} and {\it B} selected bases in which the electron spins would be measured. As in Weihs's experiment, the selections of detector settings at stations {\it A} and {\it B} were spacelike-separated from the preparation of the entangled state, and the measurements of each spin at stations {\it A} and {\it B} were spacelike-separated from each other, thereby closing the locality loophole. Meanwhile, by using an ``event-ready'' protocol, in which the joint detection of photons at station {\it C} indicated the successful preparation of an entangled state, the group could carefully monitor the total number of entangled states produced, and thereby verify that their series of spin measurements closed the fair-sampling loophole. For their first experiment, the group conducted 245 experimental runs and measured $S = 2.42 \pm 0.20$ \cite{Hanson2015}. A few months later, the group collected data on an additional 300 runs. With the combined datasets, they measured $S = 2.38 \pm 0.14$, a violation of the Bell-CHSH inequality by 2.7 standard deviations \cite{Hanson:2016}.

Soon after Hanson's group completed its first experiment, two other groups conducted experiments that likewise closed both the locality and fair-sampling loopholes \cite{Shalm:2015,Giustina:2015}. Building directly upon the 2013 experiments with high-fidelity single-photon detectors \cite{Christensen2013,Giustina2013}, each of these groups performed Bell tests with polarization-entangled photons. To close the locality loophole, the groups needed to increase the distances between the source of entangled photons and the detector stations, so that the relevant space-time events could be arranged as in the right side of Fig.~\ref{fig:locality}. For the experiment led by Krister Shalm at the National Institute of Standards and Technology (NIST) in Boulder, Colorado, the team set up a triangular arrangement. The entangled source sat at the vertex of a (nearly) right triangle; photons were transmitted to detector stations {\it A} and {\it B} via optical fibers. Stations {\it A} and {\it B} were each about 130m from the entangled source, and 185m from each other. The light travel time from the source to a detector station was thus about $0.43 \, \mu{\rm s}$. Detector settings at stations {\it A} and {\it B} were determined by QRNGs co-located at each detector, which produced fresh, random bits about every 5 ns, thereby changing multiple times while the entangled photons were in flight. The group measured clear violation of a Bell-type inequality; the probability that their experimental results could have arisen from a local-realist model, in which $p (A, B \vert {\bf a}, {\bf b})$ took the form of Eq.~(\ref{pABab}), was $p = 2.3 \times 10^{-7}$ \cite{Shalm:2015}. If one makes some simplifying assumptions and adopts Gaussian statistics, this $p$ value corresponds to a violation of the relevant inequality by more than 5 standard deviations.\footnote{\small \baselineskip 10pt As I discuss in Sec.~\ref{sec:conclusions}, there has been substantial progress and innovation in the statistical analysis of recent Bell tests, as well as in experimental designs. In particular, the use of Gaussian statistics implies several simplifying assumptions, and hence many experimental groups now report their results in terms of $p$-values rather than (or in addition to) standard deviations.}

\begin{figure}[t]
    \centering
    \includegraphics[width=3.1in]{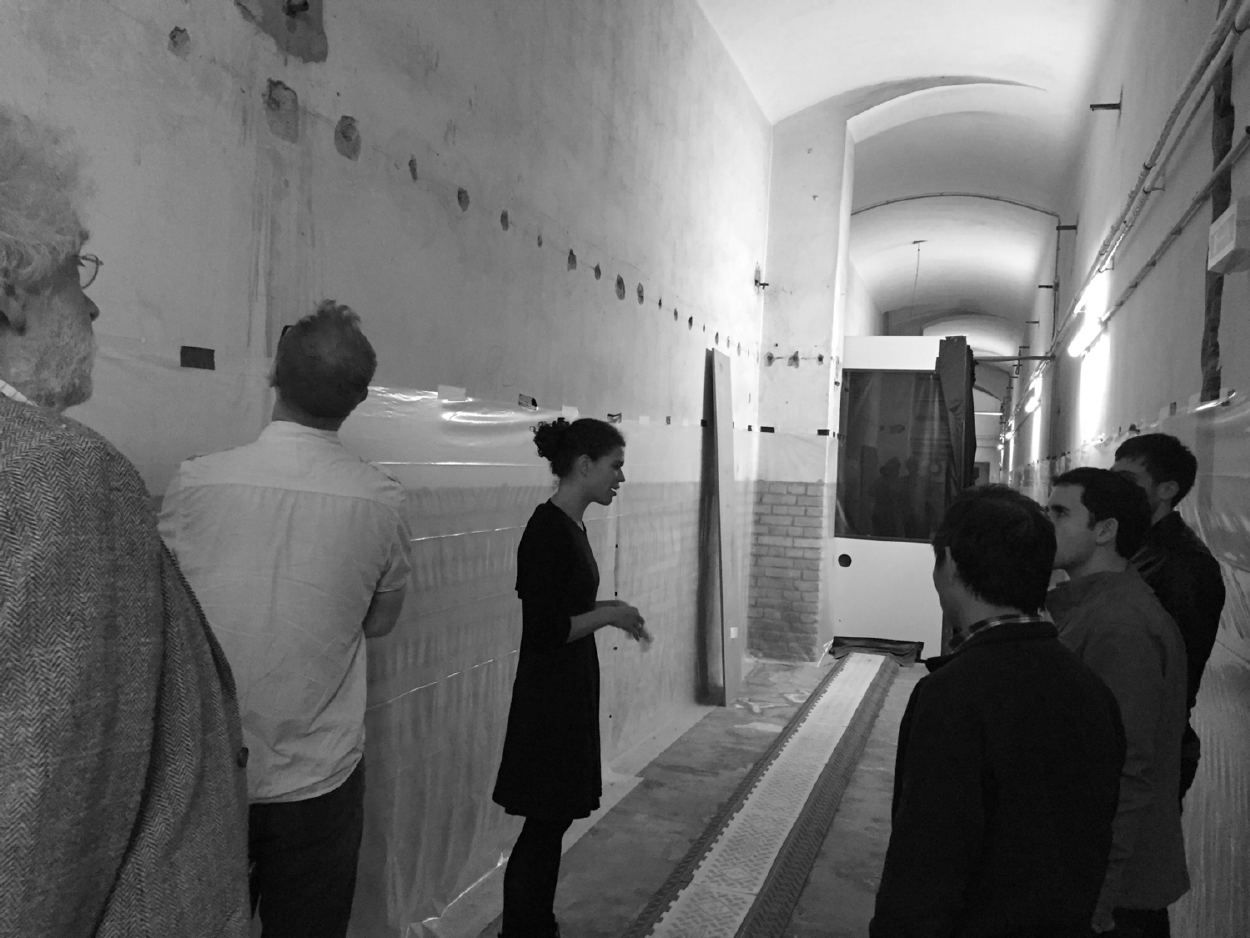}
    \caption{\small \baselineskip 10pt Marissa Giustina (center) describes her 2015 experiment to colleagues in the sub-basement of the Hofburg Palace in Vienna. The source of entangled photons is within the kiosk, visible in the middle of the hallway. Optical fibers (beneath the shielding in the center of the floor) transmitted the photons to detector stations on opposite ends of the long hallway, each 29 meters from the source. (Photo by the author.)}
    \label{fig:Giustina}
\end{figure}

At the same time, Anton Zeilinger's group in Vienna completed its own Bell test using polarization-entangled photons while closing the locality and fair-sampling loopholes. Led by Marissa Giustina, the group adopted a colinear spatial arrangement for the various experimental stations, rather than the triangular set-up of the NIST group. Optical fibers transmitted entangled photons from the central source toward detector stations on opposite ends of a long, narrow hallway in a sub-basement of the fabled Hofburg Palace in central Vienna. See Fig.~\ref{fig:Giustina}. (The project constituted a major part of Giustina's doctoral dissertation. Many graduate students {\it feel} as if they are trapped in a castle dungeon while working on their theses...) Each detector station was 29m from the source, yielding a light travel time of just $96.7$ ns from source to detector station. Within that brief time period, QRNGs at each detector station implemented fresh detector settings within 26 ns windows, ensuring spacelike-separation of the relevant events, as in the right side of Fig.~\ref{fig:locality}. Like the NIST group, the Vienna experiment measured a substantial violation of a Bell-like inequality. The probability that their measured correlations would arise in a local-realist theory of the form in Eq.~(\ref{pABab}) was just $p = 3.74 \times 10^{-31}$; if one again (naively) adopts Gaussian statistics, this result corresponds to a violation of the relevant Bell inequality by nearly 12 standard deviations \cite{Giustina:2015}. (See also Ref.~\cite{Rosenfeld:2017}.)

\section{The Freedom-of-Choice Loophole}
\label{sec:FreedomChoice}

In 1976 Shimony, Horne, and Clauser \cite{CHS:1976} identified a third significant loophole in tests of Bell's inequality, which had eluded Bell himself.\footnote{\small \baselineskip 10pt Shimony, Horne, and Clauser originally circulated their analysis \cite{CHS:1976} in the informal newsletter {\it Epistemological Letters}, which served as a forum for discussions of Bell's inequality and other issues in the foundations of quantum mechanics throughout the 1970s, at a time when many physics journals, such as the {\it Physical Review}, downplayed the topic. Several years later, their exchange with Bell was republished in the philosophical journal {\it Dialectica} \cite{LocalBeables}. Bell later included his own contributions to the exchange \cite{Bell:1976,Bell:1977} as chapters in his well-known book, {\it Speakable and Unspeakable in Quantum Mechanics} \cite{BellSpeakable}, where they appear as chapters 7 and 12. On the role of {\it Epistemological Letters}, see Ref.~\cite{FreireOpticsLab}, pp.~602-603 and Ref.~\cite{KaiserHippies}, p.~122. } (A few years later, Richard Feynman independently articulated a version of this loophole: Ref.~\cite{Feynman:1982}, p.~485.) The third loophole is often denoted the ``measurement-dependence loophole,'' the ``settings-dependence loophole,'' or the ``freedom-of-choice loophole,'' and is conceptually distinct from the locality loophole. The locality loophole concerns the flow of information {\it during} a given experimental run, and relies upon direct communication between parts of the apparatus to account for the strong correlations; locality, in other words, concerns whether $p (A, B \vert {\bf a}, {\bf b}, \lambda)$ factorizes as $p (A \vert {\bf a}, \lambda) \, p (B \vert {\bf b}, \lambda)$. The freedom-of-choice loophole, on the other hand, concerns whether any {\it common cause} could have established statistical correlations between the parameters $\lambda$ that affect measurement outcomes and the selection of detector settings $( {\bf a}, {\bf b})$, such that $p (\lambda \vert {\bf a}, {\bf b}) \neq p (\lambda)$. Such statistical correlations could arise even in the absence of direct communication between parts of the experimental apparatus. 

Shimony, Horne, and Clauser \cite{CHS:1976} pointed out that in Bell's original derivation of his inequality, he had relied upon expressions of the form in Eq.~(\ref{pABab}) for the conditional probabilities $p (A, B \vert {\bf a}, {\bf b})$. Yet the law of total probability requires that one write such expressions as 
\begin{equation}
    p (A, B \vert {\bf a}, {\bf b} ) = \int d \lambda \, p (A, B \vert  {\bf a}, {\bf b},  \lambda)  \, p (\lambda \vert {\bf a}, {\bf b} ) \, .
    \label{pABFOC}
\end{equation}
(On the law of total probability, see, e.g., Ref.~\cite{Blitzstein}, Sec.~2.3.) In general, when calculating $p (A, B \vert {\bf a}, {\bf b})$ one must take into account possible correlations between $\lambda$ and the detector settings $({\bf a}, {\bf b})$, represented by the term $p (\lambda \vert {\bf a}, {\bf b})$, {\it regardless} of whether $p (A, B \vert {\bf a}, {\bf b}, \lambda)$ factorizes as $p (A \vert {\bf a}, \lambda) \, p (B \vert {\bf b}, \lambda)$. In his original derivation, however, Bell had tacitly neglected any possible correlation between $\lambda$ and $({\bf a}, {\bf b})$, writing simply $p (\lambda)$ in place of $p (\lambda \vert {\bf a}, {\bf b})$. Via Bayes's theorem, that was equivalent to replacing $p ({\bf a}, {\bf b} \vert \lambda) \rightarrow p ({\bf a}, {\bf b})$, that is, to assuming (by fiat) that the selection of detector settings $({\bf a}, {\bf b})$ was entirely independent of the parameters $\lambda$ that could affect the behavior of the entangled particles. As Bell \cite{Bell:1976} wrote in his exchange with Shimony, Horne, and Clauser, ``It has been assumed that the settings of instruments are in some sense free variables---say at the whim of the experimenters---or in any case not determined in the overlap of the backward light cones.''  

A dozen years after Shimony, Horne, and Clauser identified the freedom-of-choice loophole, Carl Brans \cite{Brans:1988} developed an explicit local-realist model that exploited nontrivial correlations $p (\lambda \vert {\bf a}, {\bf b}) \neq p (\lambda)$ in order to mimic the predictions from quantum mechanics for Bell tests. More recent theoretical work has clarified that the freedom-of-choice loophole offers by far the most efficient means by which local-realist models can produce correlations that exceed Bell's inequality. Only a {\it minuscule} amount of statistical correlation between the selection of detector settings $({\bf a}, {\bf b})$ and the parameters $\lambda$ is required for local-realist models to mimic the correlations of a maximally entangled quantum state like $\vert \Psi^{(\pm)} \rangle$ in Eq.~(\ref{psipm}), for example---over twenty times less coordination (or ``mutual information'') than required for local-realist models that exploit the locality loophole in order to reproduce the quantum-mechanical predictions.\footnote{\small \baselineskip 10pt One may quantify the amount of correlation required for a local-realist model to mimic the quantum-mechanical predictions by exploiting the freedom-of-choice loophole in terms of the mutual information, $I = \sum_{\lambda, {\bf a}, {\bf b}} p (\lambda \vert {\bf a}, {\bf b} ) \, p ({\bf a}, {\bf b}) \, {\rm log}_2 [ p (\lambda \vert {\bf a}, {\bf b} ) / p (\lambda) ]$. The most efficient local-realist models that can reproduce predictions for correlations in a maximally entangled two-particle state by exploiting the freedom-of-choice loophole require merely $I = 0.046 \simeq 1/22$ of a bit of mutual information \cite{Friedman:2019}. Local-realist models that exploit the locality loophole in order to mimic the quantum-mechanical predictions for Bell tests, on the other hand, require at least 1 full bit of mutual information \cite{Hall:2011}. } (See Refs.~\cite{Hall:2010,Hall:2011, Hall:2015,BarrettGisin:2011,Banik:2012,Friedman:2019}.) 
In other words, despite claims that have occasionally been made in the literature, local-realist models that exploit the freedom-of-choice loophole certainly do {\it not} require the strong assumption of ``superdeterminism,'' in which experimenters' every single action would be determined by initial conditions (set, for example, at the time of the Big Bang). Rather, the freedom-of-choice loophole merely requires that in a small fraction of experimental runs, the source of entangled particles could predict (better than chance) at least one of the detector settings, ${\bf a}$ or ${\bf b}$, that would be used on a given run \cite{Friedman:2019}.

The freedom-of-choice loophole thus appears to be quite robust, theoretically: the basic rules for calculating conditional probabilities require that the term $p (\lambda \vert {\bf a}, {\bf b})$ be included in expressions like Eq.~(\ref{pABFOC}), and the possibility that $p (\lambda \vert {\bf a}, {\bf b}) \neq p (\lambda)$ offers the most efficient mechanism by which a local-realist model could yield strong correlations among measurement outcomes. So how might one address the freedom-of-choice loophole experimentally? The remarkable experiments described in Sec.~\ref{sec:FairSampling}, which managed to close both the locality and fair-sampling loopholes, relied upon quantum random number generators (QRNGs) to select detector settings $({\bf a}, {\bf b})$ for each experimental run. According to quantum mechanics, the outputs of such devices should be intrinsically random, and hence unpredictable. But the purported intrinsic randomness of quantum mechanics is part of what is at {\it stake} in tests of Bell's inequality, so the use of QRNGs in Bell tests raises the specter of circularity. Put another way: if the world were in fact governed by a local-realist theory rather than by quantum mechanics, then the behavior of QRNGs should---in principle---be susceptible to a description of the form in Eq.~(\ref{pABFOC}), including the critical term $p (\lambda \vert {\bf a}, {\bf b}).$ If that conditional probability incorporated even modest statistical correlations between $\lambda$ and the selection of various detector settings $({\bf a}, {\bf b})$, then the outputs of the QRNGs might very well {\it appear} to be random---that is, they might pass the usual suite of randomness tests, such that knowledge of the previous $N$ bits would not suffice to predict bit $N + 1$ at greater than chance levels. Yet the outputs could nonetheless have been sufficiently correlated with the (unseen) parameters $\lambda$ to produce measurement outcomes that yield $S \rightarrow S_{QM}^{\rm max} = 2 \sqrt{2}$ in tests of the Bell-CHSH inequality. (See also Ref.~\cite{Pironio:2015}.)

Physicists have pursued two distinct approaches to address the freedom-of-choice loophole in recent experiments. One approach has been to crowd-source seemingly random bits in real time. During the course of a single day---November 30, 2016---about 100,000 volunteers around the world, dubbed ``Bellsters,'' played a specially designed video game. Their task was to try to produce an unpredictable sequence of $0$'s and $1$'s; while they played, a sophisticated machine-learning algorithm analyzed each Bellster's first few entries and tried to predict what the next one would be. With real-time feedback from the algorithm, players could improve their scores by making their next selections less predictable. The outputs from all those volunteers---which totaled nearly $10^8$ (quasi-)random bits---were directed via high-speed networks to twelve laboratories distributed across five continents: from Australia to Shanghai, Vienna to Barcelona, Buenos Aires to Boulder, Colorado. In each of those laboratories that day, the real-time Bellster bits determined which detector settings $({\bf a}, {\bf b})$ would be used, run by run, in independent Bell tests. Every participating laboratory measured statistically significant violations of Bell's inequality \cite{BBT}.

A complementary approach to addressing the freedom-of-choice loophole has been to isolate the events that determine detector settings $({\bf a}, {\bf b})$ as much as possible, in space and time, from the rest of the experiment. In place of QRNGs or large numbers of Earthbound volunteers, such ``Cosmic Bell'' experiments make use of astronomical random number generators (ARNGs) for selecting detector settings: devices that perform real-time astronomical measurements of light from distant objects, and rapidly convert some aspect of those measurements into a (quasi-)random bitstream \cite{Gallicchio:2013iva}. For example, in the experiments reported in Refs.~\cite{Handsteiner:2016ulx,Leung:2017ndn,Rauch:2018rvx}, the ARNGs used dichroic filters to rapidly distinguish light from astronomical sources that was more red or more blue than some reference wavelength.

The first Cosmic Bell test was performed in Vienna in April 2016, by a collaboration including Anton Zeilinger and his group together with astrophysicists at MIT, Harvey Mudd College, and NASA's Jet Propulsion Laboratory. Polarization-entangled photons were emitted from the roof of the Institute for Quantum Optics and Quantum Information (IQOQI). One detector station, on the top floor of the Austrian National Bank (about $0.5$ km from IQOQI), included an ARNG that performed real-time measurements of light from a bright, Milky Way star that was about 600 lightyears from Earth. The other detector station, in a university building (more than 1 km from IQOQI, in the opposite direction), included its own ARNG trained on a distinct Milky Way star about 1930 lightyears from Earth \cite{Handsteiner:2016ulx}. 

In keeping with the space-time arrangement shown on the right in Fig.~\ref{fig:locality}, each ARNG implemented a fresh detector setting at its station every few microseconds, while the entangled photons were in flight, and the measurement outcomes $(A, B)$ were determined at spacelike-separated events. In addition, the causal alignment of the three experimental stations and the two astronomical sources was carefully analyzed, to ensure that the causal wave front from the stellar emission event that was intended for the ARNG at the Austrian National Bank arrived at its intended location before any information about that astronomical photon could have arrived at either the source of entangled photons or the distant detector station (and vice versa). In this way, the experiment closed the locality loophole. In addition, by selecting the detector settings on each run based on events that had occurred hundreds of years ago, quadrillions of miles from Earth, the experiment pushed back to 600 years ago the most recent time by which any local-realist mechanism could have exploited the freedom-of-choice loophole to engineer the necessary correlations between detector settings and properties of the entangled photons. The experiment measured $S = 2.502 \pm 0.042$, violating the Bell-CHSH inequality by nearly 12 standard deviations \cite{Handsteiner:2016ulx}.

\begin{figure}[t]
    \centering
    \includegraphics[width=2.1in]{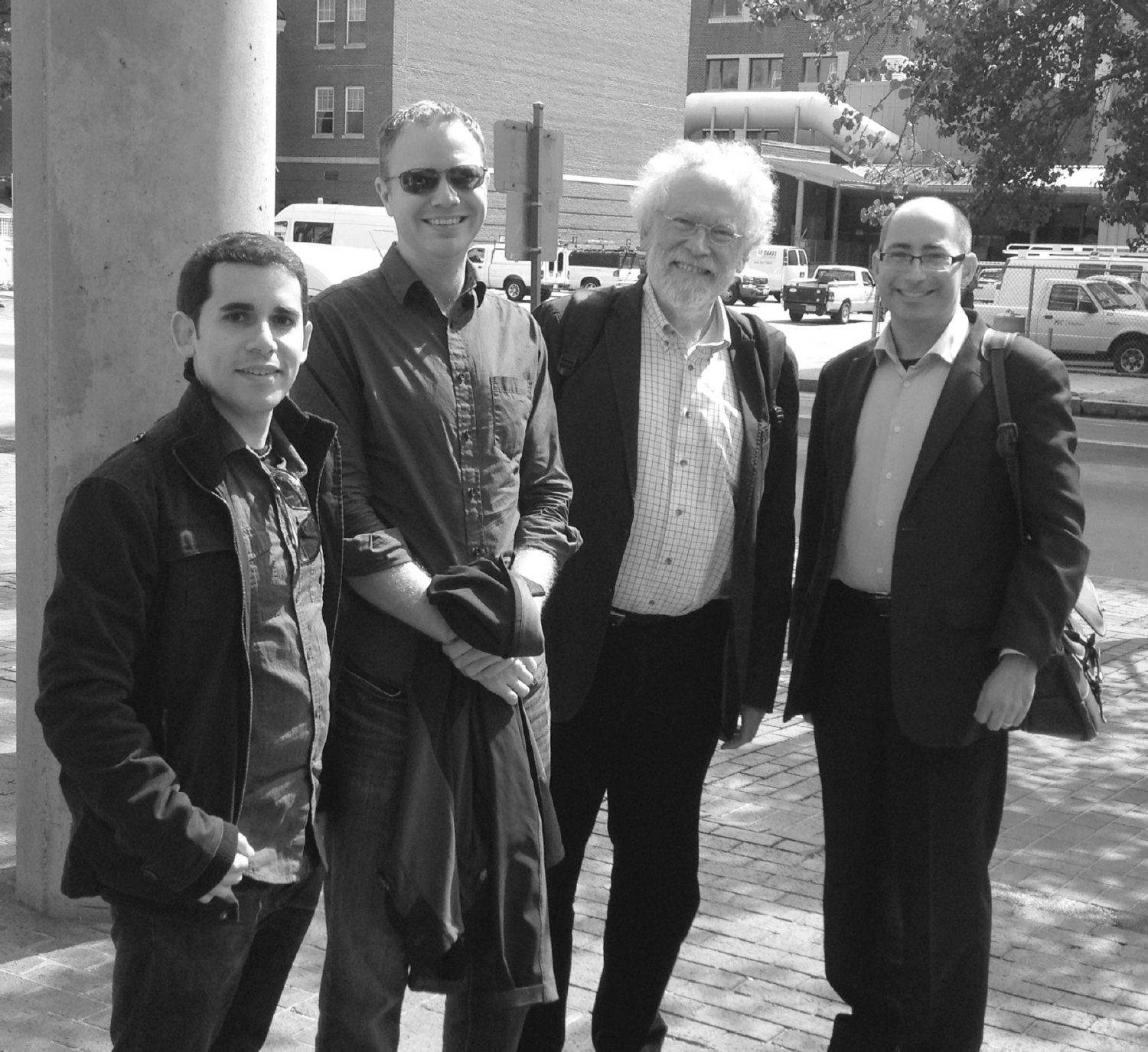}  \includegraphics[width=2.04in]{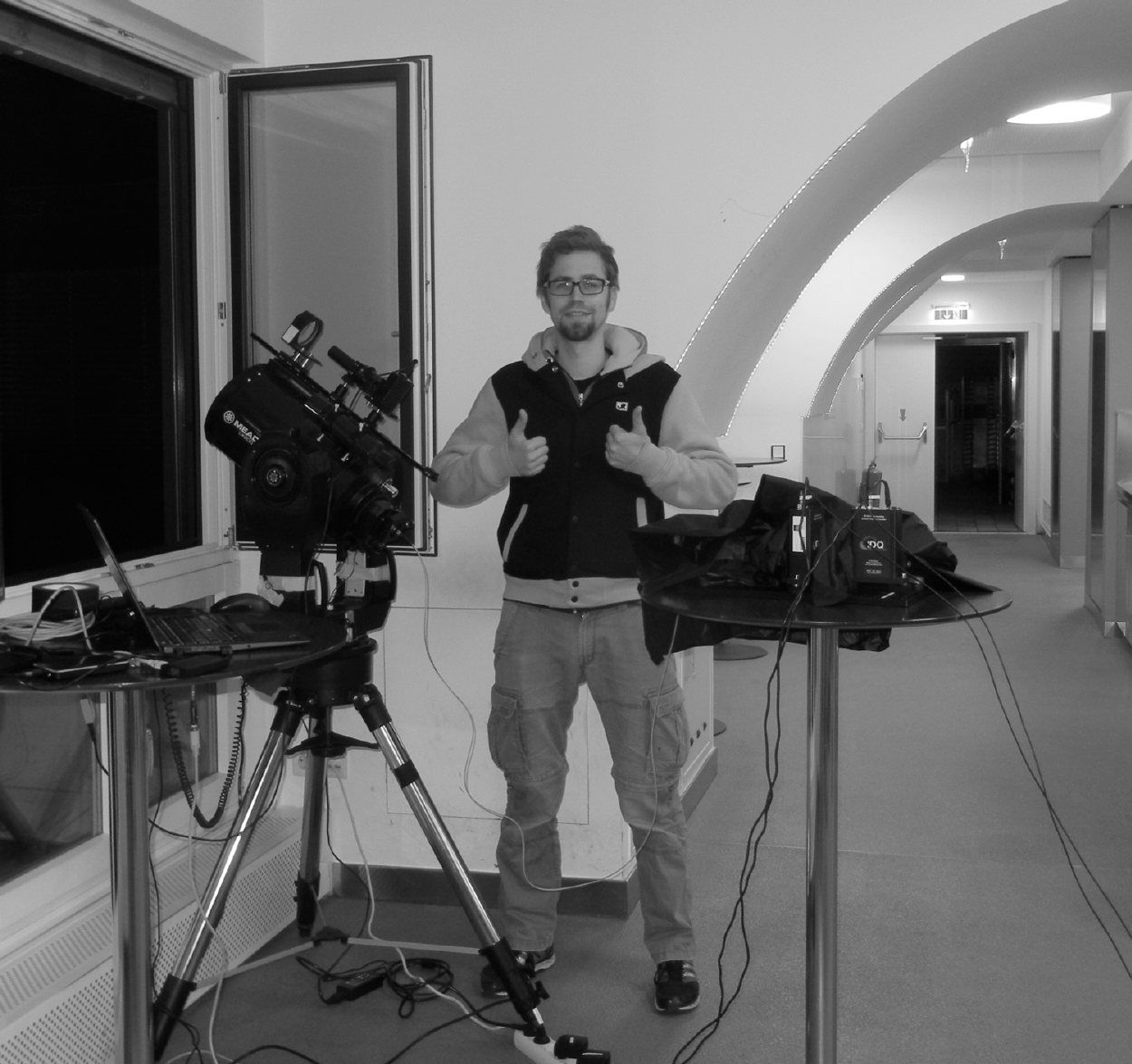} \\
    \includegraphics[width=2.1in]{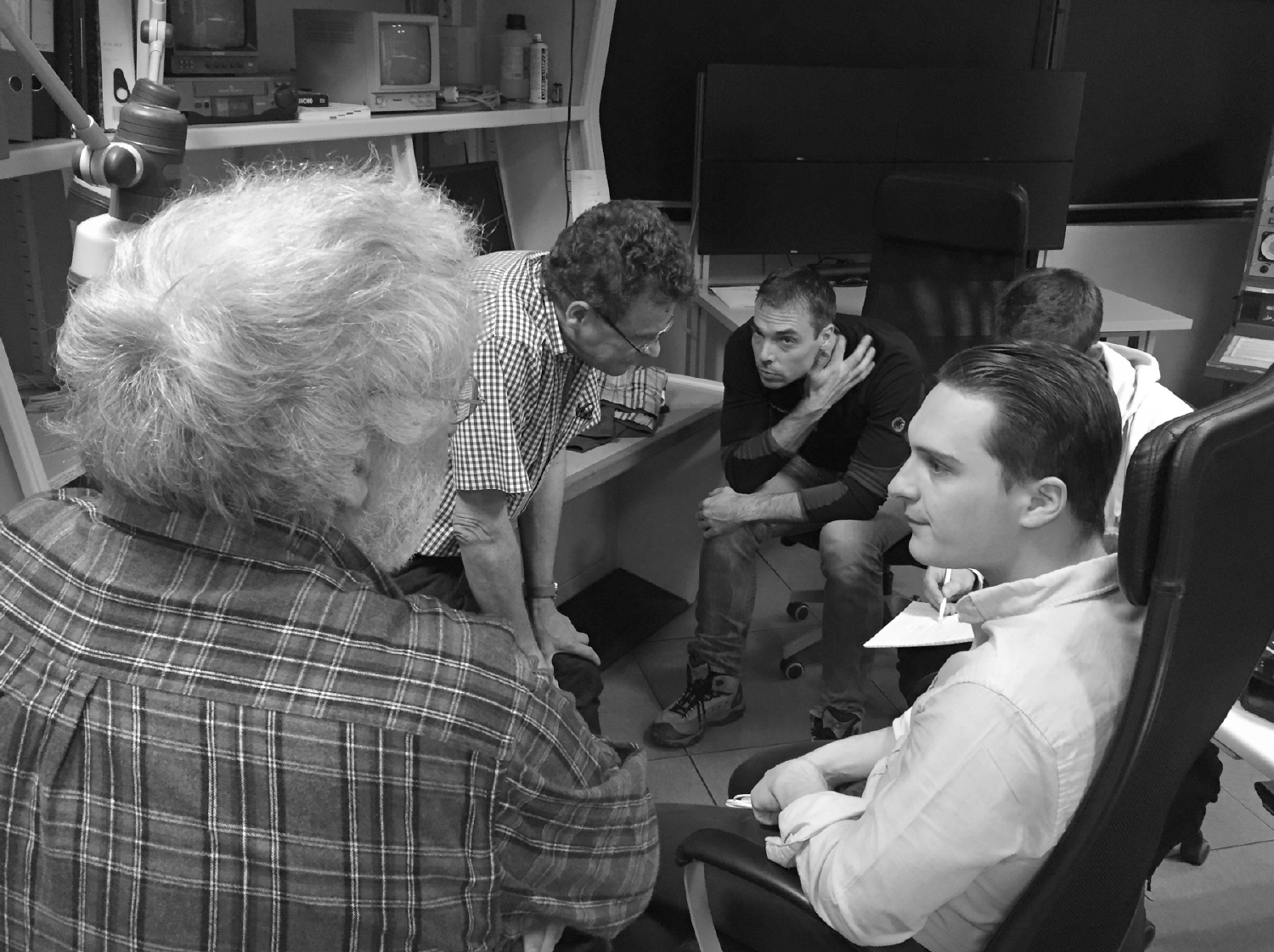} \includegraphics[width=2.1in]{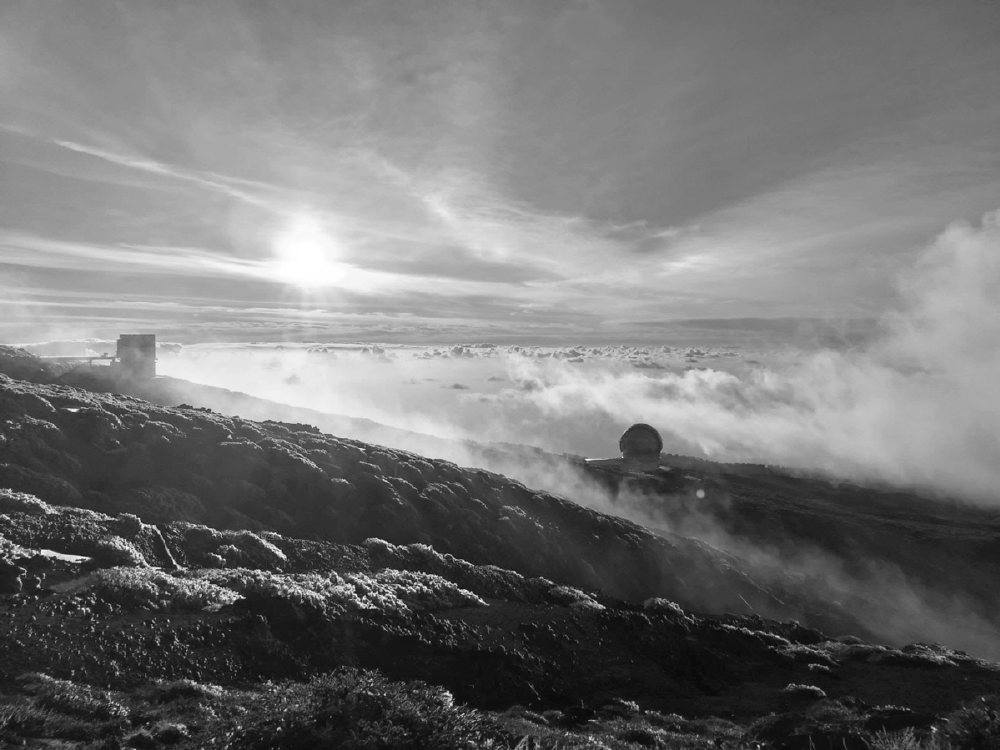}
    \caption{\small \baselineskip 10pt ({\it Top left}) Andrew Friedman, Jason Gallicchio, Anton Zeilinger, and David Kaiser discuss early plans for Cosmic Bell tests at MIT, October 2014. (From the author's collection.) ({\it Top right}) Johannes Handsteiner sets up the astronomical random number generator (ARNG) in the Austrian National Bank in Vienna for the first Cosmic Bell test, April 2016. (Courtesy S\"{o}ren Wengerovsky.) ({\it Bottom left}) Anton Zeilinger (back to the camera) discusses observing options in the control room of the William Herschel Telescope on La Palma, January 2018. Others shown (from left to right) are Christopher Benn (leaning), Thomas Scheidl, Armin Hochrainer, and Dominik Rauch. (Photo by the author.) ({\it Bottom right}) Two of the large telescopes at the Roque de los Muchachos Observatory in La Palma; on the left is the Telescopio Nazionale Galileo, which the Cosmic Bell group used during its January 2018 experiment. (Courtesy Calvin Leung.)}
    \label{fig:CosmicBell}
\end{figure}

In January 2018, the group performed a second Cosmic Bell test, this time using a pair of 4m telescopes at the Roque de los Muchachos Observatory atop La Palma, in the Canary Islands. See Fig.~\ref{fig:CosmicBell}. With the larger telescopes, the ARNGs could measure light from cosmologically distant sources: high-redshift quasars rather than Milky Way stars. The group performed measurements on pairs of polarization-entangled photons while detector settings for each experimental run were determined by emission events that had occurred 7.78 billion years ago for one detector station and 12.21 billion years ago for the other. (For reference, the Big Bang occurred $13.80$ billion years ago \cite{Planck:2018}.) As with the first Cosmic Bell experiment, causal alignment was carefully analyzed to ensure that no information about a given cosmic emission event could have arrived at either the source of entangled particles or at the distant detector before that cosmic photon was measured by its intended ARNG. Fresh detector settings were implemented within brief windows (of order $1 \, \mu$s) while the entangled photons were in flight, again closing the locality loophole. The experiment measured $S = 2.646 \pm 0.070$, violating the Bell-CHSH inequality by more than 9 standard deviations. By deploying ARNGs focused on cosmologically distant quasars, the experiment pushed back to nearly 8 billion years ago the most recent time by which any local-realist influences could have exploited the freedom-of-choice loophole to engineer the observed violation of the Bell-CHSH inequality. Given the space-time arrangement of the particular quasars used for the experiment, the past light cones of each emission event, and the expansion history of the universe since the Big Bang, this second Cosmic Bell experiment excluded such local-realist, freedom-of-choice scenarios from $96.0\%$ of the space-time volume of the past light cone of the experiment, extending from the Big Bang to the present time \cite{Rauch:2018rvx}. (See also Ref.~\cite{KaiserQL}, chap.~4.)

The two Cosmic Bell experiments thus managed to close the locality loophole while constraining the freedom-of-choice loophole by dozens of orders of magnitude, compared to earlier, pioneering efforts to address this third loophole \cite{Scheidl:2010,Aktas:2015}. Because the Cosmic Bell experiments relied upon free-space transmission of the entangled photons rather than transmitting the photons via low-loss optical fibers, however, they were not able to close the fair-sampling loophole. In a stunning accomplishment, a separate group in Shanghai led by Jian-Wei Pan conducted a Bell test in which they distributed entangled photons via optical fibers across relatively short distances: about 90m between the source and each detector station. The (vacuum) light travel time between source and each detector station was therefore about $300$ ns, requiring a correspondingly faster rate of generating and implementing fresh detector settings for each experimental run. Pan's group used ARNGs focused on very bright, nearby stars, with which they could implement fresh detector settings within windows as brief as 250 ns.\footnote{\small \baselineskip 10pt The brighter the astronomical object, the greater the flux of astronomical photons that an ARNG can collect per unit time, and hence the quicker an ARNG can output a bitstream of (quasi-)random numbers.} The group measured violations of a Bell-like inequality with $p = 7.87 \times 10^{-4}$ (roughly equivalent to 3.4 standard deviations), while closing the locality and fair-sampling loopholes and constraining any local-realist, freedom-of-choice scenario to have been set in motion no more recently than 11.5 years prior to the experiment \cite{Li:2018}. 

\section{Conclusions}
\label{sec:conclusions}

John Bell first formulated his now-famous inequality in 1964, and physicists have subjected Bell's inequality to experimental tests since 1972. Virtually every published test has measured violations of Bell's inequality, consistent with predictions from quantum mechanics. Over that period, physicists have identified several significant loopholes and devised clever, updated experimental designs, all with the goal of producing the most compelling evidence possible with which to evaluate the core question that had animated Bell's work: is the universe governed by a theory compatible with local realism, or not? Bolstered by a recent slew of experiments that have addressed various combinations of the major loopholes, \cite{Hanson2015,Hanson:2016,Shalm:2015,Giustina:2015,Rosenfeld:2017,BBT,Handsteiner:2016ulx,Rauch:2018rvx,Li:2018}, 
the evidence against local realism is stronger than ever.

Recent efforts to test Bell's inequality have featured advances in statistical analysis as well as experimental design. It was common in early experiments, for example, to adopt Gaussian statistics when analyzing the statistical significance of a measured violation of (say) the Bell-CHSH inequality. Yet Gaussian statistics rely on several simplifying assumptions. First, and most obvious, the Gaussian distribution is an idealized form that holds in the limit of an infinite number of measurements, $N \rightarrow \infty$; when applied to any finite series of measurements, one must adopt an additional assumption about the convergence of the actual statistical distribution to the idealized form. Several recent Bell tests have involved large numbers of measurements on pairs of particles, with $N \sim 10^4 - 10^7$ \cite{Christensen2013,Giustina2013,Shalm:2015,Giustina:2015,Handsteiner:2016ulx,Rauch:2018rvx,Li:2018}, which might plausibly justify the approximation $N \rightarrow \infty$, though other recent tests (especially those involving event-ready protocols) have included as few as $N\sim 10^2 - 10^3$ measurements \cite{Matsukevich:2008,Hanson2015,Hanson:2016,Rosenfeld:2017}. In other words, successful tests of Bell-type inequalities do not always approximate the domain $N \rightarrow \infty$ for which Gaussian statistics might be appropriate.

More important, use of the Gaussian distribution is predicated on the assumption that relevant variables for each experimental run are independent and identically distributed (often abbreviated as ``i.i.d.''). This assumption is usually violated in real experiments. For example, whatever processes are used to select detector settings $({\bf a}, {\bf b})$ on a given experimental run usually do not yield precisely equal numbers of trials with the various joint-settings pairs $({\bf a}, {\bf b})$, $({\bf a}', {\bf b})$, $({\bf a}, {\bf b}')$, and $({\bf a}', {\bf b}')$---hence a local-realist mechanism could exploit the ``excess predictability'' that certain combinations of detector settings can be expected to arise more frequently than others \cite{Kofler:2014}. Even more subtle, the i.i.d.~assumption neglects what have come to be called ``memory'' effects. Like a seasoned poker player who carefully tracks the cards that have been played and updates her strategy over the course of a game, a local-realist mechanism could make use of information about the previous detector settings and measurement outcomes and adjust its strategy over the course of an experiment. Exploiting such locally available information would remain compatible with local-realist conditional probabilities of the form in Eq.~(\ref{pABab}), let alone Eq.~(\ref{pABFOC}). Taking into account both excess predictabilities and ``memory'' effects therefore requires more sophisticated calculations of the statistical significance with which measured correlations in a Bell test exceed what could be accounted for by a local-realist scenario 
\cite{Kofler:2014,Gill:2003,Gill:2012,Bierhorst:2015,Elkouss:2016,Handsteiner:2016ulx,Rauch:2018rvx}.

In recent years, several groups of physicists have found an additional motivation for performing tests of Bell's inequality, beyond the enduring question about local realism. Quantum entanglement is now at the core of new devices, including quantum computation and quantum encryption. Such real-world technologies will only function as expected if entanglement, as described by quantum mechanics, is a robust fact of nature rather than an illusion that arises from some local-realist underpinning. In particular, many quantum encryption protocols rely upon embedded Bell tests to verify the security of a communication channel. If some local-realist mechanism could exploit loopholes like locality, fair sampling, and/or freedom of choice to produce the expected results in a Bell test, then (in principle) such mechanisms would also be available to eavesdroppers or hackers, intent on gaining access to unauthorized information.
(See esp.~Refs.~\cite{ekert,gisin,Scarani,AlleaumeSurvey,PanReviewLargeScale,PanReviewRealisticQKD,PirandolaReview}.) Some of the most ambitious and audacious recent tests of Bell's inequality---including the breathtaking experiment by Jian-Wei Pan and his group, involving pairs of polarization-entangled photons emitted from the specially built {\it Micius} satellite, in low-Earth orbit, and measured at detector stations 1200 km apart from each other on Earth \cite{yin1}---have been key components in building and testing real-world quantum encryption infrastructure \cite{liao,MiciusQKD}. In our new era of quantum information science, the stakes for tests of local realism have only grown, even beyond the deep questions that drove Bell, Clauser, Horne, Shimony, and their early colleagues to pursue tests of Bell's inequality.

\section*{Acknowledgements}

I am grateful to Olival Freire for inviting me to contribute to this {\it Handbook} and for many years of engaging discussion about the history of Bell's theorem, and to Guido Bacciagaluppi and Jason Gallicchio for helpful comments on an early draft of this chapter. I am also grateful to my colleagues from the ``Cosmic Bell'' collaboration, in particular Andrew Friedman, Jason Gallicchio, Alan Guth, Johannes Handsteiner, Calvin Leung, Dominik Rauch, Thomas Scheidl, and Anton Zeilinger, for years of close collaboration and camaraderie. This chapter is dedicated to the memory of Andrew S.~Friedman (1979-2020). This work was supported in part by the U.S. National Science Foundation INSPIRE grant No.~PHY-1541160. Part of this work was conducted in MIT's Center for Theoretical Physics, and supported in part by the U.S. Department of Energy under Contract No.~DE-SC0012567.



%

\end{document}